\documentclass{article}
\usepackage[margin=1.0in]{geometry}
\usepackage{jcappub}
\usepackage{graphicx}
\usepackage{amsmath}
\usepackage{amssymb}
\usepackage{hyperref}
\usepackage{natbib}
\usepackage{multirow}
\hypersetup{colorlinks,linkcolor=blue,citecolor=blue}

\title{Tests of Neutrino and Dark Radiation Models from Galaxy and CMB surveys}

\author[a]{Arka Banerjee,}
\author[b]{Bhuvnesh Jain,}
\author[a]{Neal Dalal,}
\author[a]{and Jessie Shelton}

\affiliation[a]{Department of Physics, \\University of Illinois at Urbana-Champaign, \\1110 West Green Street, Urbana, IL 61801-3080 USA}
\affiliation[b]{Department of Physics and Astronomy,\\ Center for Particle Cosmology, \\University of Pennsylvania, \\Philadelphia, PA 19104 USA}

\emailAdd{abanerj6@illinois.edu}
\emailAdd{bjain@physics.upenn.edu}
\emailAdd{dalaln@illinois.edu}
\emailAdd{sheltonj@illinois.edu}
\abstract{
We analyze the ability of galaxy and CMB lensing surveys to constrain massive neutrinos and new models of dark radiation. We present a Fisher forecast analysis for neutrino mass constraints with the LSST galaxy survey and the CMB S4 survey. A joint analysis of the three galaxy and shear 2-point functions from LSST, along with key systematics parameters and Planck priors, can constrain the neutrino masses to $\sum m_\nu = 0.041\,$eV at 1-$\sigma$ level, comparable to constraints expected from Stage 4 CMB lensing. If low redshift information from upcoming spectroscopic surveys like DESI is included, the constraint becomes $\sum m_\nu = 0.032\,$eV. These constraints are derived having marginalized over the number of relativistic species ($N_{\rm eff}$), which is somewhat degenerate with the neutrino mass.
We also explore the gain by combining LSST and CMB S4, that is, using the five relevant auto- and cross-correlations of the two datasets. We conclude that advances in modeling the nonlinear regime and the measurements of other parameters are required to ensure a neutrino mass detection. Using the same datasets, we explore the ability of LSST-era surveys to test ``nonstandard'' models with dark radiation. We find that if evidence for dark radiation is found from $N_{\rm eff}$ measurements, the mass of the dark radiation candidate can be measured at a 1-$\sigma$ level of $0.162\,$eV for fermionic dark radiation, and $0.137\,$eV for bosonic dark radiation, for $\Delta N_{\rm eff} = 0.15$.
We also find that the NNaturalness model of Arkani-Hamed et al \cite{Arkani-Hamed2016}, with extra light degrees of freedom, has a sub-percent effect on the power spectrum: even more ambitious surveys than the ones considered here will be needed to test such models.
}
\begin{document}
\maketitle 
\flushbottom

\section{Introduction}

Over the last few decades, flavor oscillation experiments e.g.~\cite{SuperK98,SNO2001,K2K2003} have established that neutrinos in the Standard Model are massive, and have three mass eigenstates. These experiments have accurately measured two of the mass splittings as well as the mixing angle between the various eigenstates. However, these experiments do not directly measure the absolute masses of the neutrinos, or the mass hierarchy of the three species. 

The presence of massive neutrinos has non-trivial consequences for cosmology. The fraction of the total energy density of the universe that is contributed by neutrinos is proportional to the sum of the masses of the three neutrino species \citep{Dodelson2003book}. For most realistic neutrino masses, all three mass eigenstates were relativistic prior to last scattering of the Cosmic Microwave Background (CMB), but at least two mass eigenstates are non-relativistic today. These neutrinos contribute to the overall matter density at late times, but with  little clustering below their free streaming scale. For light neutrinos, the free streaming scale can be comparable to the Hubble radius. This is unlike the Cold Dark Matter (CDM) component which clusters strongly on all scales at low redshifts to form halos and filaments. The clustering of neutrinos has the effect of damping the growth of the matter power spectrum on small scales when compared to a CDM-only case \cite{Eisenstein1997}.

Since the damping of the power spectrum depends on the neutrino energy density, and therefore on the sum of the neutrino masses, an accurate measurement of the matter power spectrum can provide a strong constraint on the sum of the neutrino masses. Combined with the two mass splittings that have been measured from terrestrial experiments, this would allow for an accurate determination of the masses of individual eigenstates and the hierarchy.

Apart from neutrinos in the Standard Model, many theories of physics beyond the Standard Model include light, weakly interacting degrees of freedom in their particle content. While the details of these particles' interactions may be very different, as long as the interaction rates are small on cosmological time scales, the effect of these light particles on the observable matter power spectrum will be similar to that of the Standard Model neutrinos. These extra light species will damp the power spectrum on scales below their free streaming scale by an amount proportional to the energy density in that species.

Recently many models of dark sectors with varied particle content and interactions have been proposed, \cite{Spergel1999,CyrRacine2012,Fan2013,Ackerman:mha,Feng:2009mn,Tulin:2013teo,Carlson:1992fn,Hochberg:2014dra,Kopp:2016yji,Dror:2016rxc,Forestell:2016qhc,Pappadopulo:2016pkp,Kuflik:2015isi,Agashe:2014yua,Loeb:2010gj,Chacko:2015noa} for example.  Many of these models predict the existence of some form of thermal dark radiation, both bosonic and fermionic. Pseudo-Goldstone bosons are the most common example of the former, and are realized in many extensions of the Standard Model. Among fermionic dark radiation candidates, sterile neutrinos are perhaps the best-motivated and extensively studied. In all these cases, the presence of extra dark radiation in the universe would lead to a non-zero $\Delta N_{\rm eff}$ in measurements of the CMB, i.e.\ a change to the effective number of relativistic degrees of freedom at CMB last scattering. However, since these particles are expected to be relativistic at the epoch of CMB, a measurement of $\Delta N_{\rm eff}$ alone does not measure the masses of these particles. As long as the masses of these particles are light enough that they are fully non-relativistic today, the damping of the matter power spectrum is directly proportional to the mass of the dark radiation particle, and measurements of this damping on small scales can potentially constrain the mass, both for fermions and for bosons \cite{Hannestad:2005}. 

One specific example of such a model with extra light particles is the ``NNaturalness'' mechanism \cite{Arkani-Hamed2016} proposed to solve the hierarchy problem. This model introduces $N$ non-interacting copies of the Standard Model field content, with the Standard Model being identified as the copy with the lowest non-zero Higgs vacuum expectation value. One of the predictions of this model is that massive neutrinos from sectors close to our Standard Model could have energy densities not too much smaller than the energy density of the Standard Model neutrinos. Apart from a signature on $N_{\rm eff}$ at CMB, there will also be extra damping of the low redshift matter power spectrum compared to the Standard Model, which is potentially observable.

Accurate measurement of the late time matter power spectrum, therefore, will be extremely important for constraining the neutrino mass in the Standard Model, as well as for constraining light degrees of freedom from more exotic models. One probe of late-time clustering of mass is lensing of the CMB by intervening matter. The {\it Planck} experiment along with ground based experiments such as SPT \cite{SPT} and ACT \cite {ACT}, have already produced lensing maps of the CMB, and future planned experiments such as the Simons Observatory and CMB Stage 4 will be able to this more accurately and down to smaller scales \cite{S4ScienceBook,SIMONS}. 

%Various authors  e.g.~\cite{Lesgourgues2005,Kaplinghat2003,Allison2015}  have looked into  
%how well these experiments will be able to constrain the neutrino mass.

Another powerful method for determining the late time matter power spectrum is weak lensing measurements in large galaxy photometric surveys, like the ongoing Dark Energy Survey \cite{DES2005}, Subaru HSC survey \cite{HSC} and KiDS \cite{KIDS}, and the upcoming surveys by the LSST \cite{LSST}, Euclid \cite{EUCLID} and WFIRST \cite{WFIRST} missions..  While measurements of galaxy-galaxy lensing and galaxy autocorrelations from these surveys do not individually measure matter auto-correlations, due to degeneracy with the unknown galaxy bias, the combination of the two on linear scales (where the bias is expected to be deterministic) eliminates the bias uncertainty \cite{Baldauf2016}.  Inclusion of cosmic shear measurements from these surveys further enhances the ability to pin down the underlying matter power spectrum. Using weak lensing from current and upcoming CMB and photometric galaxy surveys, as well as information about low redshifts from ongoing and future spectroscopic surveys, numerous authors have investigated the bounds placed by these measurements on the neutrino mass e.g.~\cite{Hu1997,Abazajian2002,Lesgourgues2005,Kaplinghat2003,Hannestad2006,Kitching2008,Allison2015,Takada2008, Takada2009, Takada2011, Abazajian2016, Dunkley2016, Verde2009, Carbone2011, Verde2015, Verde2010, Hamann2012, DESI, S4ScienceBook, Wu2014,Abazajian2013,Zhao2013,Carbone2012,Gratton2007,Wang2005,Takeuchi2013,Font-Ribera2013,Archidiacono2016,Hall2012,dePutter2014,dePutter2009,Lahav2010,Mueller2014}. With the inclusion of smaller scales in successive generations of cosmological experiments, as well as lower experimental noise, these bounds have gotten progressively tighter, to the point that current data from the {\it Planck} experiment offers stronger bounds on the sum of neutrino masses than any terrestrial experiment, albeit with assumptions about the background cosmology.

%, like the Dark Energy Survey \cite{DES2005} and the upcoming LSST survey \cite{LSST}.

In this paper, we present a Fisher forecast analysis of the ability of a survey like LSST to constrain the sum of the neutrino masses, as well as its ability to constrain other beyond Standard Model light degrees of freedom using weak lensing measurements. In \S \ref{method}, we present our method for calculating constraints on cosmological parameters, as well for differentiating between different models producing changes in the angular power spectrum $C_l$. In \S \ref{systematics}, we overview the survey parameters for LSST and CMB Stage 4 lensing experiment, as well as parameterize the systematic uncertainties in these surveys. In \S \ref{constraints}, we present the constraints on the sum of neutrino masses from LSST and compare it to those that will be obtained from CMB Stage 4 lensing. We also present constraints from the joint analysis of the two. In \S \ref{darkrad}, we discuss how observables at LSST can provide constraints on dark radiation. In \S \ref{NNat}, we discuss the prospects of detecting models of NNaturalness using galaxy clustering and lensing at LSST. Finally, in \S \ref{summary}, we summarize our findings, and discuss future avenues of study.

 \section{Method}
 \label{method}
 \subsection{Weak lensing in galaxy surveys}
 \label{method_lensing}
 Galaxy surveys like LSST provide maps of galaxy distributions and shear on the sky, which can be used to construct the different 2-point correlation functions that will be used in our analysis: the galaxy-galaxy autocorrelation function $C_l^{gg}$, the galaxy-convergence cross spectrum $C_l^{g\kappa}$ and the convergence autospectrum $C_l^{\kappa\kappa}$. Here we make use of the simple relation between shear spectra and convergence spectra.  Using the Limber approximation these spectra are constructed from the underlying 3-dimensional power spectra \cite{Hu2004}:
 \begin{equation}
  C_l^{x_ix_j} = \int dz \frac{H}{D_A^2}W_i(z)W_j(z)P^{s_is_j}(k = l/D_A;z) \, ,
 \end{equation}
 where $x_i$ stand for $g$ or $\kappa$, while $s_i$ stand for the underlying 3 dimensional source fields. $D_A$ is the angular diameter distance, and $W_i$ are the weighting functions in redshift space. For the galaxy number fluctuations, the three dimensional source field is the fluctuations in the 3-dimensional number density:
 \begin{equation}
  s (\mathbf r; z) = \frac{\delta n_V}{\bar n_V} \, .
 \end{equation}
  The weighting function for galaxy fluctuations is given by 
  \begin{equation}
   W_g(z) = \frac{D_A^2}{H} \frac{\bar n _V}{\bar n_A} \, ,
  \end{equation}
 where the normalization factor $\bar n_A$ is chosen so that $\int W_g(z) dz = 1$. For the shear field the 3-dimensional source field is the fluctuation of the matter density:
 \begin{equation}
  s(\mathbf r;z) = \frac{\delta \rho_m}{\rho_m} \, ,
 \end{equation}
 and the weighting function is 
 \begin{equation}
  W_\kappa (z) = \frac 3 2 \Omega_m \frac {H_0} H \frac{H_0 D_{OL}} a \int_z ^\infty dz'\frac{D_{LS}}{D_{OS}}W_g(z') \, ,
 \end{equation}
 where $D_{OL}$ stands for the angular diameter distance to the lens, $D_{OS}$ is the angular diameter distance to the source, and $D_{LS}$ is the distance between the lens and the source.
 We use the publicly available CAMB code \cite{CAMB} to generate the various power spectra that go into our analysis.
  
  To construct the covariance matrix for the $C_l$, we assume that the different $l$
  are uncorrelated and so the covariance matrix is diagonal in $l$. We also assume
  that up to $l_{\rm max}$, we are roughly in the linear regime, where  
  Gaussian statistics are valid, and all $n$-point functions can be broken 
  down into products of various two point functions given by the $C_l$. To take 
  into account the shape noise in the shear spectra and the shot noise in the galaxy
  spectra we define 
  \begin{equation}
   \tilde C_l^{x_ix_j} = C_l^{x_ix_j} + N_l^{x_ix_j}  \, ,
  \end{equation}
  where $x$ stands for $g$ or $\kappa$ and $i$ is used to label different redshift 
  bins. The noise terms are given by 
  \begin{eqnarray}
   \label {shotnoise} N_l^{g_ig_j} &=& \frac{\delta_{ij}}{\bar n_i}\, , \\
   N_l^{\kappa_i\kappa_j} &=& \frac{\delta_{ij} \gamma_{\rm rms}^2}{\bar n_i} \, ,\\
   N_l^{g_i\kappa_j} &=& 0 \, .
  \end{eqnarray}
  Here $\bar n_i$ represent the number counts of lens and source galaxies in each redshift 
  bin in units of $sr^{-1}$. For the shape noise contribution, we use a value of $\sqrt{\gamma_{\rm rms}^2} = 0.22$.

  In the linear regime 
  all covariances of the power spectra can be written as products 
  of the power spectra themselves. Therefore the different elements of the covariance matrix 
  $\mathcal {C}_l$ can be written as:
  \begin{equation}
   \left[\mathcal{C}_l\right]^{ij,kl} \equiv \tilde C_l^{x_ix_k} \tilde C_l^{x_jx_l} 
   + \tilde C_l^{x_ix_l}\tilde C_l^{x_jx_k} \, .
  \end{equation}
  Using this covariance matrix, we construct different components of the LSST 
  Fisher matrix
  \begin{equation}
   F^{\rm LSST}_{\alpha\beta} = f_{\rm sky}\sum_l (2l+1) \sum_{ijkl}\frac{\partial C_l^{x_ix_j}}
   {\partial p_\alpha} \left[\mathcal{C}_l\right]^{-1}_{ij,kl}
   \frac{\partial C_l^{x_kx_l}} {\partial p_\beta} \, ,
  \end{equation}
  where $p_\alpha$ represent the model parameters, and $f_{\rm sky}$ is the fraction 
  of the sky covered in the survey, which we take to be 0.5 for LSST. 
  
  To use information on the parameters from {\it Planck}, we use the covariance matrices available in the 
  {\it Planck} Legacy Archive for the baseline model. We ensure that we do not include {\it Planck} constraints coming from CMB lensing within the {\it Planck} experiment.
  The total Fisher matrix is obtained by adding together the resultant {\it Planck} Fisher 
  matrix to the Fisher matrix we derive for the LSST experiment
  \begin{equation}
   F = F^{\rm LSST} + F^{Planck} \, .
  \end{equation}

  The constraint on parameter $\alpha$ is then given by 
  \begin{equation}
   \sigma (p_\alpha) = \sqrt{(F^{-1})_{\alpha\alpha}} \, .
  \end{equation}
  
  \subsection{CMB lensing}
  \label{method_cmb}
  For the CMB lensing analysis, we consider the lensing power spectrum $C_l^{dd}$ where $\mathbf d $, the deflection field, is the gradient of the lensing potential $\mathbf d = \nabla \phi$. The noise level $N_l^{dd}$ for a given set of experimental sensitivities can be estimated following \cite{Hu2002}, and was calculated using QUICKLENS \citep{quicklens}. The covariance matrix for the CMB lensing power spectrum is then 
  \begin{equation}
   \mathcal C_l = 2 \left(C_l^{dd} + N_l^{dd}\right)^2 \,, 
  \end{equation}
  and the elements of the Fisher matrix are given by 
  \begin{equation}
   F^{\rm CMB}_{\alpha\beta} = f_{\rm sky} \sum (2l+1) \frac{\partial C_l^{dd}}{\partial p_\alpha} \mathcal C_l^{-1} \frac{\partial C_l^{dd}}{\partial p_\beta} \,.
  \end{equation}
  Once again, we use {\it Planck} priors, where lensing information from {\it Planck} is not used. This is done by adding together the two Fisher matrices, as in \S \ref{method_lensing}.

  \subsection{Model differentiation}
  \label{modeldiff}
  To calculate the statistical significance of distinguishing between two models which are not connected by a parameter which can be varied smoothly, we use the following procedure \cite{Hezaveh2013}. Let the difference in the various predicted power spectra from the two models be denoted by $\delta C_l^{x_ix_j}$. The $\chi^2$ difference between the two models is then given by 
  \begin{equation}
   \chi^2 = f_{\rm sky} \sum_l (2l+1) \delta C_l^{x_ix_j}\left[\mathcal C\right]^{-1}_{ij,kl}\delta C_l^{x_kx_l} \,.
  \end{equation}
  A part of this $\chi^2$ difference can be taken into account by varying the continuous parameters in the fiducial model. The change of the observable power spectra given a change in the parameters of the fiducial model is $\delta C_l^{x_ix_j} = \left(\partial C_l^{x_ix_j}/\partial p_\alpha\right)\delta p_\beta$. Taking this variation into account, the new $\chi^2$ is given by 
  \begin{equation}
  \label{chi2_min}
   \chi^2 = f_{\rm sky} \sum_l (2l+1)\left[\delta C_l^{x_ix_j} + \delta p_\alpha \frac{\partial C_l^{x_ix_j}}{\partial p_\alpha}\right]\left[\mathcal C\right]^{-1}_{ij,kl}\left[\delta C_l^{x_kx_l} + \delta p_\beta \frac{\partial C_l^{x_kx_l}}{\partial p_\beta}\right] \, .
  \end{equation}
 By setting $\partial \chi^2/\partial p_\alpha = 0$ in the above equation, we calculate the $\delta p_\alpha$ which needs to made about the fiducial set of parameters $p_\alpha$ to obtain the minimum $\chi^2$:
 \begin{equation}
  \delta p_\alpha = - \left(F\right)^{-1}_{\alpha\beta}\frac{\partial C_l^{x_ix_j}}{\partial p_\beta}\left[\mathcal C\right]^{-1}_{ij,kl}\delta C_l^{x_kx_l} \,,
 \end{equation}
 where $F$ is the Fisher matrix of the fiducial model. The final $\chi^2$ is then calculated using \eqref{chi2_min}.  

  \section{Survey parameters and systematics}
  \label{systematics}
  \subsection{LSST}
  
  \begin{figure}
   \centering
   \includegraphics[scale = 0.4]{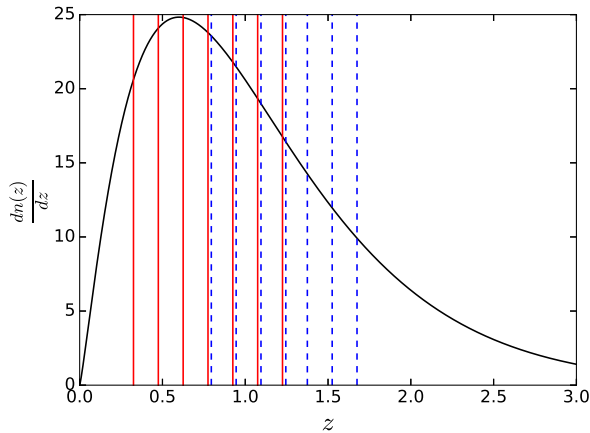}
   \caption{Expected source galaxy counts (per square arcmin)
   as a function of redshift in the LSST survey is plotted with the black curve. The solid red lines indicate the photometric redshifts $z_{\rm min}$ and $z_{\rm max}$ of the 6 lens bins used in our analysis. The dashed blue lines represent the $z_{\rm min}$ and $z_{\rm max}$ for the 6 source bins. The blue lines have been slightly displaced for clarity.}
   \label{redshift_distribution}
  \end{figure}

  To get the number counts of 
 sources in an LSST like survey, we use a redshift distribution of source galaxies given by the following form:
 \begin{equation}
  \frac{dn}{dz} \propto z^{1.2} \exp \left(- \frac{z}{0.5}\right) \,,
 \end{equation}
  with a total number density $n_{\rm source} = 30\,$arcmin$^{-2}$.
  For lens galaxies, we will use redMaGiC type of galaxies \citep{Rozo2016} which have a redshift distribution $dn/dz \propto \chi(z)^2/H(z)$, and total number density of lenses $n_{\rm lens} = 0.25\,$arcmin$^{-2}$.  
  
  In our calculations, we consider different redshift binnings - we start from 1 lens redshift bin and 1 source redshift bin, and go up to 6 lens redshift bins and 6 source redshift bins. We show the bin centers and widths for this last case in Fig. \ref{redshift_distribution}. For each lens redshift bin center, we calculate  $k_{\rm max}$ such that $\frac{k^3P(k)}{2\pi^2}\big |_{k_{\rm max}}
  \approx 0.2$. Using this, we obtain the highest multipole $l_{\rm max} = k_{\rm max} \chi$, which we use in our analysis. This is done to ensure that that even for the highest multiploles in each bin, we are in the regime where perturbations can still be treated as being roughly linear, and assumptions of the independence of different $l$ modes and linear biasing are valid. As expected, $l_{\rm max}$ increases with redshift, so we can go out to smaller scales at higher redshifts. This allows for tighter constraints on the different cosmological parameters, which affect the shape of the power spectrum, along with the overall amplitude. We tabulate the $l_{\rm max}$ for each of our bins in Table \ref{lmax_table}.

  \begin{table}
  \begin{center}
  \begin{tabular}{ |c|c|c| }
\hline
 & $z_{\rm center}$ & $l_{\rm max}$ \\ \hline
\multirow{6}{*}{Lenses} & 0.40 & 210 \\
 & 0.55 & 240 \\
 & 0.70 & 330 \\
 & 0.85 & 440 \\ 
 & 1.00 & 570 \\
 & 1.15 & 720 \\\hline

%\multirow{6}{*}{Sources} & 0.85 & 200 \\
% & 1.00 & 220 \\
% & 1.15 & 250 \\
% & 1.30 & 280 \\ 
% & 1.45 & 320 \\
% & 1.60 & 370 \\
\end{tabular}
\quad
\begin{tabular}{ |c|c|c| }
\hline
 & $z_{\rm center}$ & $l_{\rm max}$ \\ \hline
%\multirow{6}{*}{Lenses} & 0.40 & 210 \\
% & 0.55 & 240 \\
% & 0.70 & 330 \\
% & 0.85 & 440 \\ 
% & 1.00 & 570 \\
% & 1.15 & 720 \\\hline

\multirow{6}{*}{Sources} & 0.85 & 200 \\
 & 1.00 & 220 \\
 & 1.15 & 250 \\
 & 1.30 & 280 \\ 
 & 1.45 & 320 \\
 & 1.60 & 370 \\ \hline
\end{tabular}

\caption{List of redshift bin centers and $l_{\rm max}$ used for each redshift bin for the LSST survey.}
\label{lmax_table}
\end{center}
\end{table}

  \begin{figure}
%\begin{tabular}{ccc}
  \includegraphics[scale = 0.4]{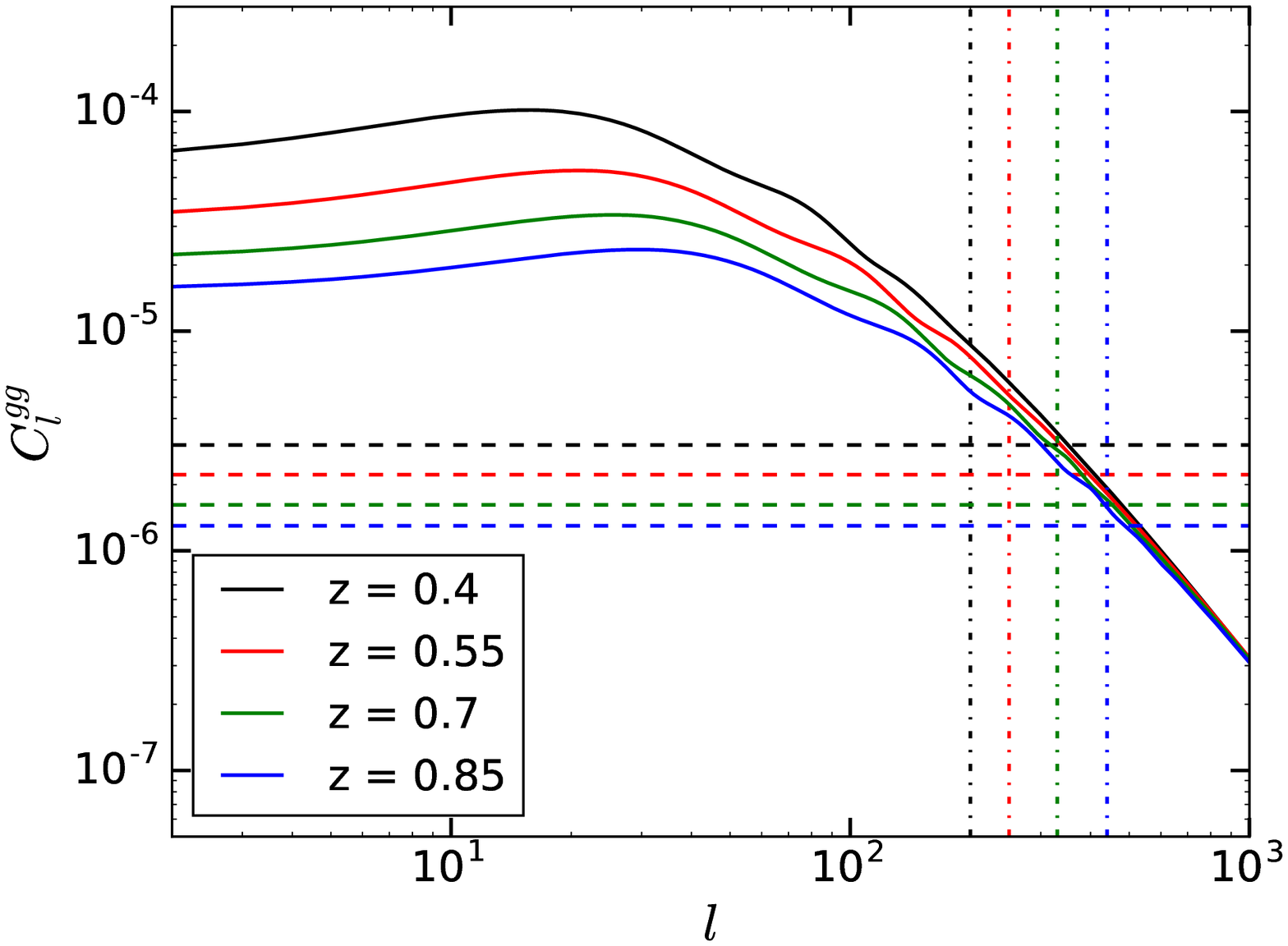}   \includegraphics[scale = 0.4]{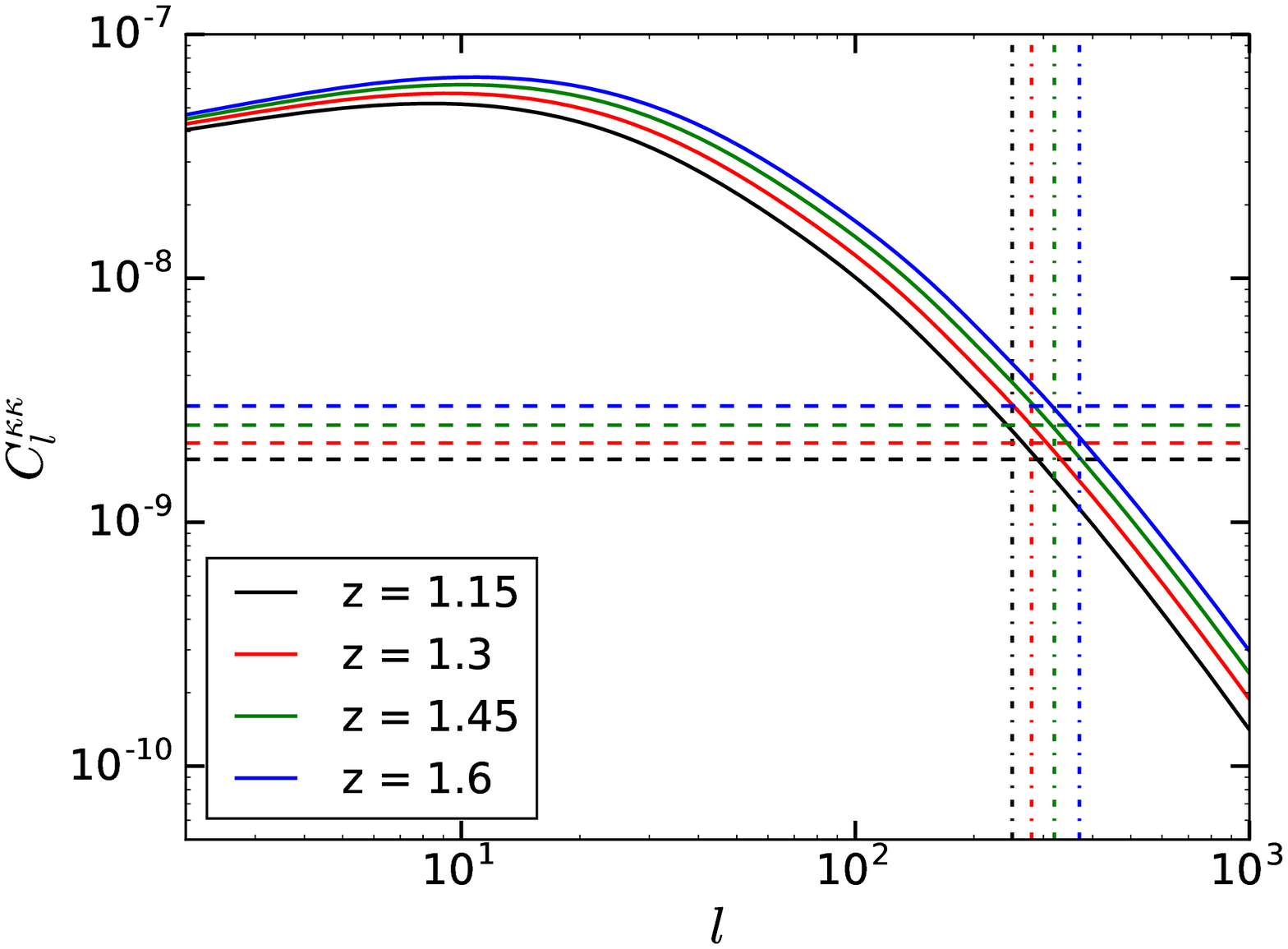}  
     %\end{tabular}
   \caption{Comparison of the $C_l^{gg}$ and $C_l^{\kappa\kappa}$ power spectra at different redshifts to the shot noise and shape noise levels at those redshifts. The solid lines plot the power spectrum, while the dashed line of the same color plots the shot noise or shape noise for that power spectrum at that redshift expected at LSST. The dot-dashed lines represent the value of $l_{\rm max}$ for each redshift bin. The spectra are sample variance dominated out to $l\sim 400$.
   }
   \label{surveynoise}
   \end{figure}

  We also note that we use the linear matter power spectrum for all our calculations. Since we restrict ourselves to the scales for which $\frac{k^3P(k)}{2\pi^2}\big |_{k_{\rm max}}
  = 0.2$, using the linear power spectrum, as opposed to the full nonlinear power spectrum is a valid choice, except for some scales near $k_{\rm max}$ and consequently $l_{\rm max}$. The differences are expected to be small, and using the linear power spectrum will always yield a conservative estimate on the constraints. 
  
  In Fig.\ \ref{surveynoise}, we plot the power spectra and the associated noise levels for the assumed number counts of lenses and sources in some of the redshift bins that we use in our calculations. In the left panel of Fig.\ \ref{surveynoise}, we plot the galaxy-galaxy autospectra $C_l^{gg}$ for four of the lens redshift bins, with the solid lines. We represent the shot noise level in each bin using the dashed lines of the same color. Similarly, in the right panel, we plot the shear autospectra $C_l^{\kappa\kappa}$ for four of the source redshift using solid lines. We also indicate the shape noise for these bins using dashed lines of the same color.

   As mentioned, the lens galaxy counts we use throughout our analysis are chosen to approximate the distribution of redMaGiC galaxies \cite{Rozo2016}. The galaxy-galaxy autorcorrelations for the lens bins are also calculated using the same subsample of galaxies. Since these galaxies form only a fraction of all the galaxies in those redshift bins, the level of shot noise in our measurements, as expressed in Eq.\ \ref{shotnoise}, will be higher than the case where all galaxies are used. However, as  Fig.\ \ref{surveynoise} shows, shot noise is sub-dominant compared to the signal covariance over the range of scales we consider, meaning that the increased shot noise from the redMaGiC subset should not degrade the errors significantly.  The advantage of using this subset is that the redshifts of redMaGiC galaxies may be determined to a very high level of accuracy using photometry alone, allowing us to effectively ignore lens redshift errors as a source of systematics. For the lens redshift bins, therefore, we only consider the galaxy bias of each bin $b_i$ as sources of systematics, and include these as nuisance parameters in our Fisher matrix analysis.

   For the source galaxies, however, we use the entire galaxy population from those redshift bins. While using these high number densities helps reduce the shape noise, we need to consider multiple sources of systematics \citep{Kwan2016}. Amongst these, the most significant systematic that we need to account for is the photometric redshift uncertainties of the source galaxies, as these are not as well measured as those in the redMaGiC sample. Another important systematic that we need to consider is the shear calibration. Here we allow for a multiplicative error arising from calibration errors in the shear and photo-z. We do not attempt to model additive, scale dependent uncertainties in the shear or clustering as a reasonable analytical model or even level of uncertainty is not available. To account for these sources of systematic errors, we introduce the nuisance parameters, $m_i$, one for each source bin. We assume that these effects can be parameterized by allowing for an overall rescaling of the shear measurements: $\kappa_i \rightarrow \kappa_i(1+m_i)$. We will consider this parameterization for the source uncertainties throughout the paper. Note that the relation between our $m_i$ and photo-z bias is not linear, and is redshift dependent. But it is a reasonably good approximation to capture both shear and photo-z bias into a multiplicative parameter for our forecasting purposes.
   
   Since we are using a single nuisance parameter $m_i$, per bin, to account for both shear calibration and redshift uncertainties, as the calibration of the shear in LSST gets better, the $m_i$ will mostly encode our uncertainty on the photometric redshifts of the source galaxies.
   
   %We note that our current analysis ignores certain other sources of systematic errors. For example, we have not included errors arising from intrinsic alignments of source galaxies, known to affect lensing measurements in galaxy surveys e.g.~\cite{Troxel2014}. We also neglect systematic errors sourced by alignments of lens and source galaxies \cite{Hirata2004}. These systematics have been been explored in the context of LSST in \cite{Schaan2016}. Another source of systematics which is not explored in this work are the additive errors in the shear measurements due to residuals from the Point Spread Function (PSF) correction \cite{Chang2013}.  \red{Don't we need some text explaining why it's OK to neglect all of these?}
   
   We note that our analysis assumes that other sources of systematic errors are subdominant to statistical errors and the systematic uncertainty due to calibration errors that we have modeled. These include intrinsic alignments of galaxies \cite{Hirata2004,Troxel2014}, and additive errors in the shear measurements due to residuals from the Point Spread Function (PSF) correction \cite{Chang2013}.  For intrinsic alignments (in particular the GI alignments), a recent study \cite{Schaan2016} has shown that the systematic uncertainty is well below statistical errors for an analysis similar to ours (see also \cite{Krause2015,Joachimi2010,Troxel2014,Hall2014}). Additive errors in the shear are not generally modeled in forecast studies as they are exceedingly difficult to anticipate and various mitigation strategies are employed to deal with them in the measurement and analysis.
   
   \subsection{CMB lensing}
   
\begin{figure}
%\begin{tabular}{ccc}
\centering
  \includegraphics[scale = 0.4]{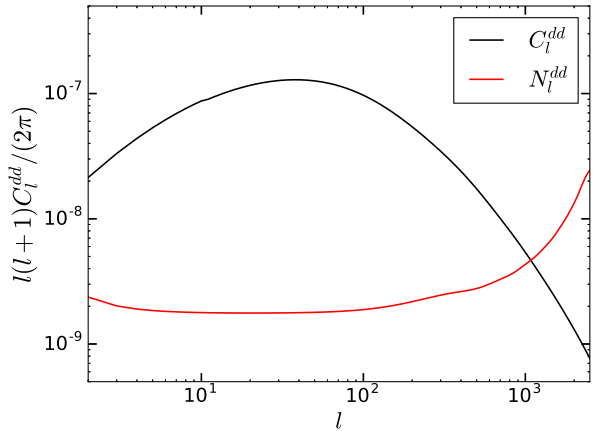}    
     %\end{tabular}
   \caption{Noise level $N_l^{dd}$ in the deflection power spectrum $C_l^{dd}$ from CMB Stage 4 lensing for the assumed survey parameters. The lensing signal is sample variance dominated out to $l\sim 1000$.
   }
   \label{cmbnoise}
   \end{figure}

   A number of specifications have been proposed for the survey parameters of CMB Stage 4 experiments, in this paper, we will use the following specifications. We consider fractional sky coverages $f_{\rm sky} = 0.75$, $f_{\rm sky}=0.5$ and $f_{\rm sky} = 0.25$.  The beam size is assumed to be $1'$ in all cases and we assume an overall experimental sensitivity of $0.58$ $\mu$K-arcmin. In Fig.\ \ref {cmbnoise}, we plot the deflection power spectrum $C_l^{dd}$, and the noise level $N_l^{dd}$ on this observable for the survey parameters that we assume. 
   
   For CMB lensing we include information from multipoles up to $l_{\rm max} = 3000$, while using a lower cutoff $l_{\rm min} = 30$. Since the lensing kernel for CMB peaks at around $z=2$, it is justified for us to go up to this high $l_{\rm max}$ while still using assumptions of linearity for the power spectra and their covariances.
   
  \section{Neutrino mass constraints}
  \label{constraints}
  We assume a fiducial model with  
  $\left(\tau,n_s,\ln[10^{10}A_s],\sum m_\nu,N_{\rm eff},\Omega_m,\Omega_b,\Omega_\Lambda,h,w,b_i,m_i\right)
  = $ \newline 
  $(0.066,0.967,3.15,0.06,3.046,0.3,0.05,0.7,0.7,-1,1.5,0)$ where $b_i$ stands for the bias in different redshift bins - with one bias parameter per redshift bin. While there may be additional signatures in the galaxy bias \cite{Loverde2014, Loverde2016}, we do not attempt to model them or use the additional information they may provide - our forecasts are therefore conservative. The bias parameters will be marginalized over when deriving constraints on various cosmological parameters like the sum of the neutrino masses $\sum m_\nu$ and the dark energy equation of state $w$. In our analysis, we do not assume any priors on the bias parameters - the constraints on these parameters come from data only. Apart from the bias parameters, we also marginalize over the shear uncertainties $m_i$ defined in \S \ \ref{systematics}, and whose fiducial value we assume to be $0$. Our fiducial constraints on $\sum m_\nu$ and $w$ are derived without assuming priors on the nuisance parameters $m_i$. We discuss the effect of placing priors on these parameters later in this section. We note that in all our calculations, we have assumed $\Omega_k = 0$, that is, curvature is neglected. Further, we have assumed that the equation of state of dark energy $w$ is time invariant.
  
  %\begin{table}
 %\begin{center}
 %  \begin{tabular}{|c|c|c|c|}
 % \hline
 % $N_L$ & $N_S$  & $\sigma\left(\sum m_\nu\right)$  (eV)  & $\sigma (w)$\\
 % \hline
 % 1 & 1 & 0.093 & 0.069\\
 % 4 & 4 & 0.052 & 0.028\\
 % 6 & 6 & 0.041 & 0.020\\
 % \hline
 %\end{tabular}
 %\caption{1-$\sigma$ constraints on the neutrino mass, and the dark energy equation of state $w$ at LSST for different number of lens and source bins used in the analysis.}
 %\label{masstable}
 %\end{center}
 %\end{table}
 
 \subsection{LSST constraints on neutrino mass}
   
\begin{figure}
%\begin{tabular}{ccc}
  \centering
  \includegraphics[scale = 0.35]{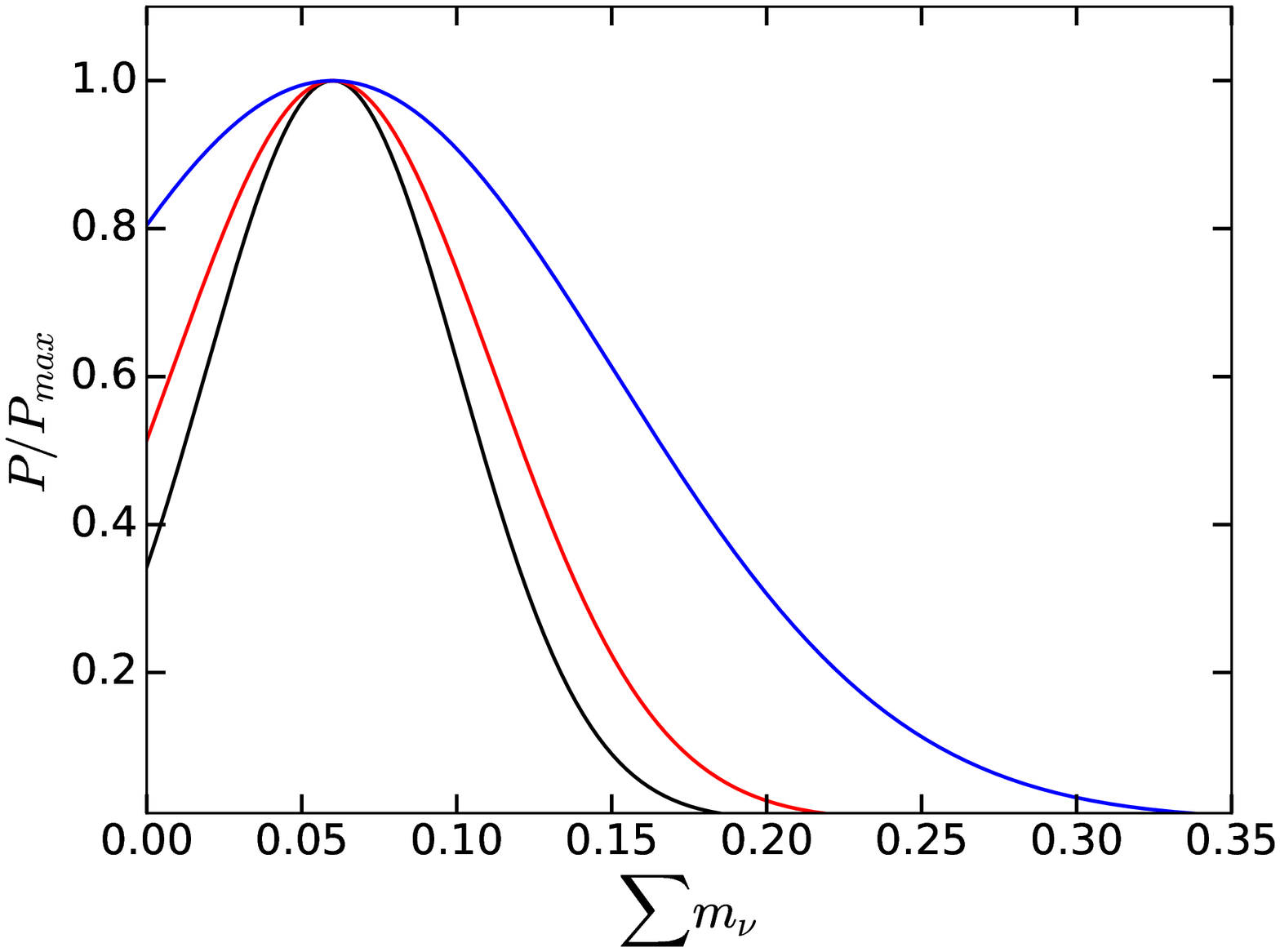}    \includegraphics[scale = 0.35]{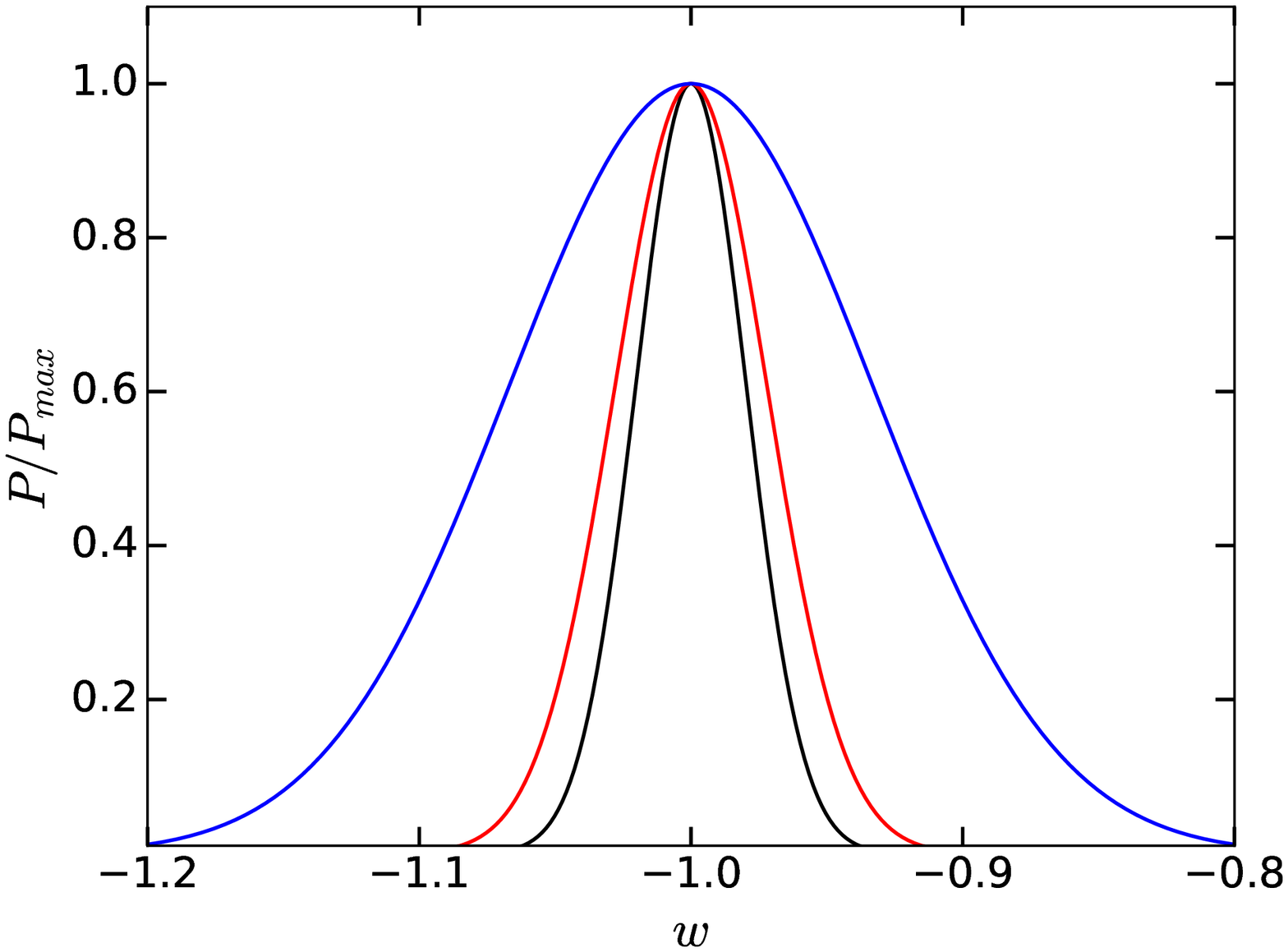}\\
     %\end{tabular}
   \caption{Marginalized probability distributions for the sum of neutrino masses (left), and dark energy equation of state $w$ (right), from LSST.
   The blue, red and black curves correspond to $N_L = N_S = 1,4,6$ respectively. Increasing the number of redshift helps extract more tomographic information, but this gain saturates as the individual bins become too thin.}
   \label{1dprob}
   \end{figure}

\begin{figure}
  \centering
%\begin{tabular}{ccc}
   \includegraphics[scale = 0.4]{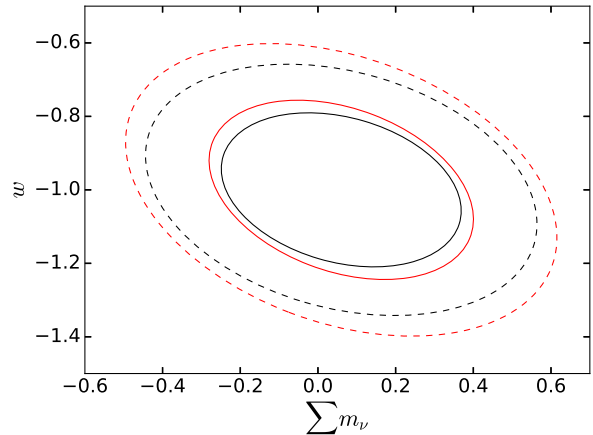} \\
     %\end{tabular}
   \caption{1-$\sigma$ (solid lines) and 2-$\sigma$ (dashed lines) confidence intervals on 2-d subspace of the parameters $\sum m_\nu$ and $w$ from LSST. The red curves are the results from 4 source bin and 4 lens bin analysis. The black curves represent the results when 6 source bins and 6 lens bins are used.}
   \label{2dprob}
   \end{figure}

  We use the formalism described in \S \ \ref{method_lensing} to obtain constraints on cosmological parameters for different number of lens and source bins at LSST. We find that for bins of fixed width, increasing the number of bins improves the constraints on the neutrino mass. This happens for two reasons - by increasing the number of bins, we increase the redshift coverage of the sources and lenses. Secondly, using more bins allows for the measurement of higher number of cross-correlations. This improvement in the constraints is summarized in Table \ref{masstable}, and is illustrated in Fig. \ref{1dprob}, where we plot the marginalized probability distributions for $\sum m_\nu$ and $w$ for different number of bins used in the analysis. In Fig \ref{2dprob} we plot the 1-$\sigma$ and 2-$\sigma$ confidence intervals for pair of the parameters $\left(\sum m_\nu,w\right)$ for two choices of source and lens bin numbers - $4$ and $6$. While the gain in going from one source and one lens bin to 4 of each gives a large improvement in the constraints, the constraints start to saturate as we add more bins. This is because the finer binning leads to smaller galaxy counts in each bin, raising the shot noise level.  Using 6 lens and 6 source bins yields a constraint on the sum of the neutrino masses $\sigma(\sum m_\nu) = 0.041\,$eV. From the same analysis, we obtain a constraint of $0.020$ on the dark energy equation of state $w$.

   \begin{table}
 \begin{center}
   \begin{tabular}{|c|c|c|}
  \hline
     & $\sigma\left(\sum m_\nu\right)$  (eV)  & $\sigma (w)$\\
  \hline
  $N_S=1$, $N_L = 1$ & 0.093 & 0.069\\
  $N_S=4$, $N_L = 4$ & 0.052 & 0.028\\
  $N_S=6$, $N_L = 6$& 0.041 & 0.020\\
  $N_S=6$, $N_L = 6$ (+ DESI\cite{DESI})& 0.032 & 0.017\\
  $N_S=6$, $N_L = 6$ (+ DESI\cite{DESI} + $N_{\rm eff}$ prior))& 0.028 & 0.016\\
  \hline
 \end{tabular}
 \caption{Forecasts of 1-$\sigma$ constraints on the neutrino mass, and the dark energy equation of state $w$ at LSST for different number of lens and source bins, and different priors. The $N_{\rm eff}$ prior used was $0.03$, following \cite{S4ScienceBook}. }
 \label{masstable}
 \end{center}
 \end{table}

    Since we do not assume any priors on the biases of the lens galaxies, we check how well the biases are constrained by data. We find that the bias parameters in each lens redshift bin is constrained at about $2\%$, with the bias parameters for the low redshift lens bins being slightly better constrained than the bias parameters for the higher redshift bins. This happens because of two opposing effects. The lower redshift bins have a higher signal to noise ratios in their galaxy clustering power spectra. On the other hand, we go out to a higher $l_{\rm max}$ for the higher redshift, increasing the sensitivity of those redshift bins to the bias parameters. These two effects roughly cancel each other out to provide similar constraints on the bias parameters for all redshift bins we consider.
 
 %\begin{figure}
 % \centering
 %\includegraphics[scale = 0.4]{biasprob.eps}
 %\caption{1-$\sigma$ constraints on the bias from different lens bins. $b_1$ is the bias of the lens bin with the lowest redshift and $b_6$ is the bias of the lens bin with the highest redshift. The bias parameters for lower redshift bins are slightly better constrained than those for the higher redshift bins.}
 %\label{bias_distribution}
%\end{figure}

Apart from LSST, other future cosmological surveys will also be sensitive to the effects of neutrino mass on galaxy clustering and weak lensing observables. It is, therefore, useful to check how the constraints on the cosmological parameters depend on survey specifications, such as the sky coverage, redshift depth, and number densities of galaxies. Apart from the LSST survey specifications, we consider two other surveys - one with specifications similar to EUCLID \cite{EUCLID}, and the other with survey parameters similar to the WFIRST survey \cite{WFIRST}. For the EUCLID-like survey we assume $f_{\rm sky} = 0.375$, an average source redshift $z=0.7$, and number density $n_{\rm source} =20\,\rm{arcmin}^{-2}$. The 1-$\sigma$ constraint on $\sum m_\nu$ for this survey is $0.060\,$eV. For the WFIRST-like mission, we assume $f_{\rm sky} = 0.0675$, an average source redshift $z = 1.4$, and number density $n_{\rm source} =40\,\rm{arcmin}^{-2}$, and we find a 1-$\sigma$ constraint of $0.067\,$eV on $\sum m_\nu$. Note that the constraints for WFIRST are not very different from the other surveys even though it has small $f_{\rm sky}$ due to the greater survey depth. Therefore, if WFIRST continues beyond its nominal three year mission, the sky coverage $f_{\rm sky}$ could become larger, and the constraints will improve accordingly. We find that our constraint forecasts are weaker compared to a similar forecast for the EUCLID mission in \cite{Hamann2012} in the case where the authors assume that the linear galaxy bias is known exactly for the lens galaxies. When this assumption is relaxed, the authors find a similar constraint on the neutrino mass as the ones presented in this paper. For the gain by combining all three surveys, see also \cite{Jain2015}.

To account for systematic errors in our constraint forecasts at LSST coming from the shear bias of the source redshift bins, along with the photometric redshift uncertainties, we introduce the nuisance parameters $m_i$ for each source redshift bin, as mentioned in \S \ \ref{systematics}. As with the bias parameters, these are marginalized over to yield constraints on the cosmological parameters of interest. Note that photometric redshift uncertainty leads to biased estimates of the distances to galaxies, which are not exactly degenerate with shear calibration. But it is sufficiently accurate for the purposes of this paper to fold the two biases into a single bias parameter per redshift bin. 

Unlike with the galaxy bias parameters, which must be measured from the same dataset used for cosmology, we test the effect of priors on these shear bias parameters as they can be estimated with image simulations or high resolution imaging. We then study how the constraints on the sum of neutrino masses and the equation of state for dark energy vary with the imposed prior. We find that when flat priors are imposed on the $m_i$, as is the case for our fiducial results, the constraint on the neutrino mass from LSST weak lensing obtained above degrades by $\sim 17\%$, whereas the constraint on $w$ degrades by only $\sim 5\%$, compared to the case where these parameters are completely ignored. 
   
\begin{figure}
 \centering
 \includegraphics[scale = 0.5]{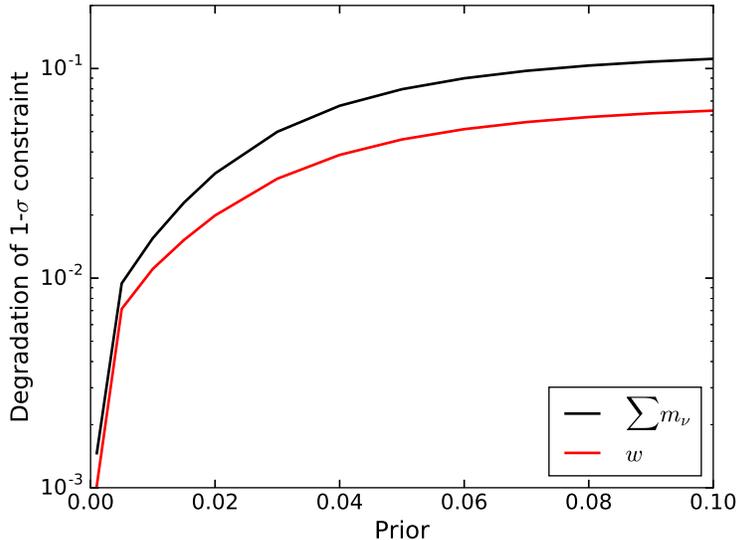}
 \caption{Degradation of the 1-$\sigma$ constraints on the parameters as a function of the prior assumed on $m_i$, the shear calibration of the source bins at LSST. The constraints change very weakly with the value of the prior, especially for $w$.}
 \label{error_degradation}
\end{figure}

   We illustrate how the constraints on the cosmological parameters change as we impose stronger priors in Fig. \ref{error_degradation}. For an imposed prior of $0.02$, which may be achievable in LSST, we find that the neutrino mass constraint degraded by only about $3\%$ compared to the case where the effect of these parameters is ignored. The degradation on the constraint on $w$ is once again, even smaller - roughly $1\%$. Obtaining percent-level priors on shear bias would require extensive and realistic image simulations, as planned  by LSST, and/or deep, high resolution imaging of a large enough subset of the source galaxies -- e.g. with space based imaging by WFIRST. An alternative is to actually use CMB lensing to calibrate the shear -- this approach was investigated by \cite{Das2009} and more recently in some detail for a joint analysis of LSST and CMB Stage 4 lensing by \cite{Schaan2016}. They find a calibration of shear bias in LSST at the percent level is indeed feasible. Note however that this still leaves redshift bias as a systematic, so for our fiducial results we use flat priors on the $m_i$ parameters. 
   
%%BJ: the para below shouild probably be modfied: it's only a small part that comes from galaxy clustering alone. See a partial attempt at editing below. 
   The behavior of the constraints as seen in Fig. \ref{error_degradation} suggests that significant self-calibration is possible even with no external priors. This is due to several factors: the combination of auto- and cross-spectra across redshift bins gives $N(N+1)/2$ shear spectra for $N$ redshift bins, while there are only $N$ shear bias parameters. Additional self-calibration comes from galaxy-shear cross-spectra which have a different scaling with redshift, and some information from galaxy-galaxy autospectra alone, which are not affected by the shear bias and photometric redshift errors of the source redshift bins. Even though the galaxy-galaxy autospectra depend on the \textit{a priori} unknown galaxy bias of the lenses, the comparatively small error bars that will be achievable in LSST, coming mainly from the large sky coverage $f_{\rm sky}$, means that these autospectra are also sensitive to parameters which modify the shape of the various $C_l$. Massive neutrinos change both the amplitude and the shape of the 3 dimensional matter power spectrum, along with the different $C_l$. Changes in the dark energy equation of state changes the cosmological growth factor, thereby changing the amplitude of the various $C_l^{gg}$. In addition, the dark energy equation of state also affects the distance-redshift relation, which in turn, affects the positions of the BAO peaks in $C_l^{gg}$. Since this second effect is not degenerate with a change in the different bias parameters, the shapes of $C_l^{gg}$ from different redshift bins provides strong constraints on $w$.
%{\bf \textcolor{red} {Constraints with DESI BAO priors to be added.}}

%\begin{table}
% \begin{center}
%   \begin{tabular}{|c|c|c|}
%  \hline
%  Survey  & $\sigma\left(\sum m_\nu\right)$ (eV)  & $\sigma (w)$\\
%  \hline
%  LSST & 0.041 & 0.020\\
%  Euclid  & 0.060 & 0.031\\
%  WFIRST & 0.067  & 0.036\\
%  \hline	
% \end{tabular}
% \caption{1-$\sigma$ constraints on the neutrino mass, and the dark energy equation of state $w$ for different surveys.}
% \label{surveytable}
% \end{center}
% \end{table}

%\begin{figure}
% \centering
 %\includegraphics[scale = 0.5]{nuprob_survey.eps}
% \includegraphics[scale = 0.35]{nuprob_survey.eps}    \includegraphics[scale = 0.35]{wprob_survey.eps}
% \caption{Comparison of the marginalized probability distribution for $\sum m_\nu$ and $w$ from different surveys. The solid lines represent constraints from the LSST survey, the dashed lines are for a survey with specifications like the Euclid survey. The dot-dashed lines represent constraints from a WFIRST-like survey.}
% \label{surveycomparison}
%\end{figure}
  \subsection{Comparison to CMB Stage 4 lensing}
\begin{figure}
%\begin{tabular}{ccc}
\centering
  \includegraphics[scale = 0.35]{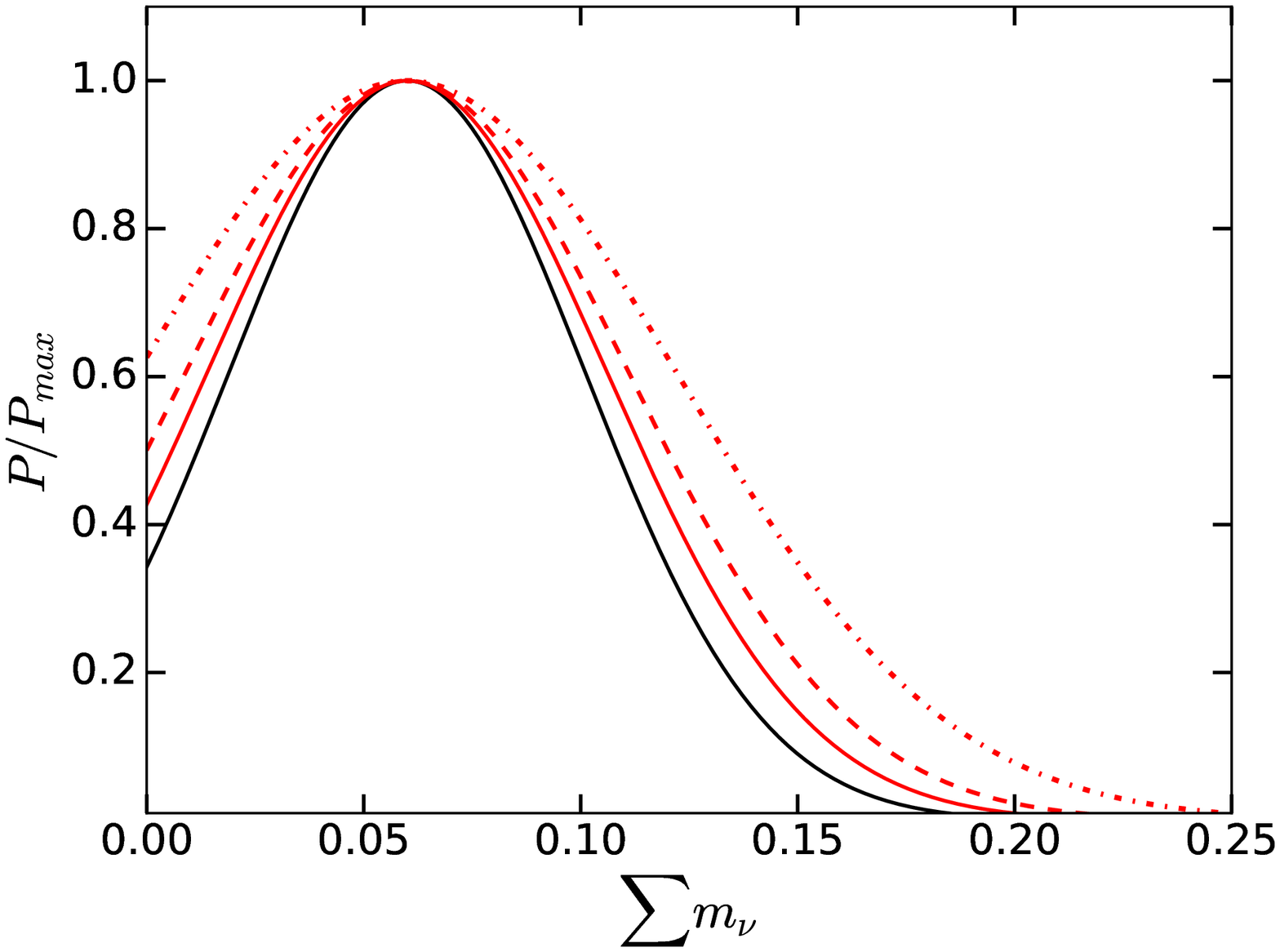}     \includegraphics[scale = 0.35]{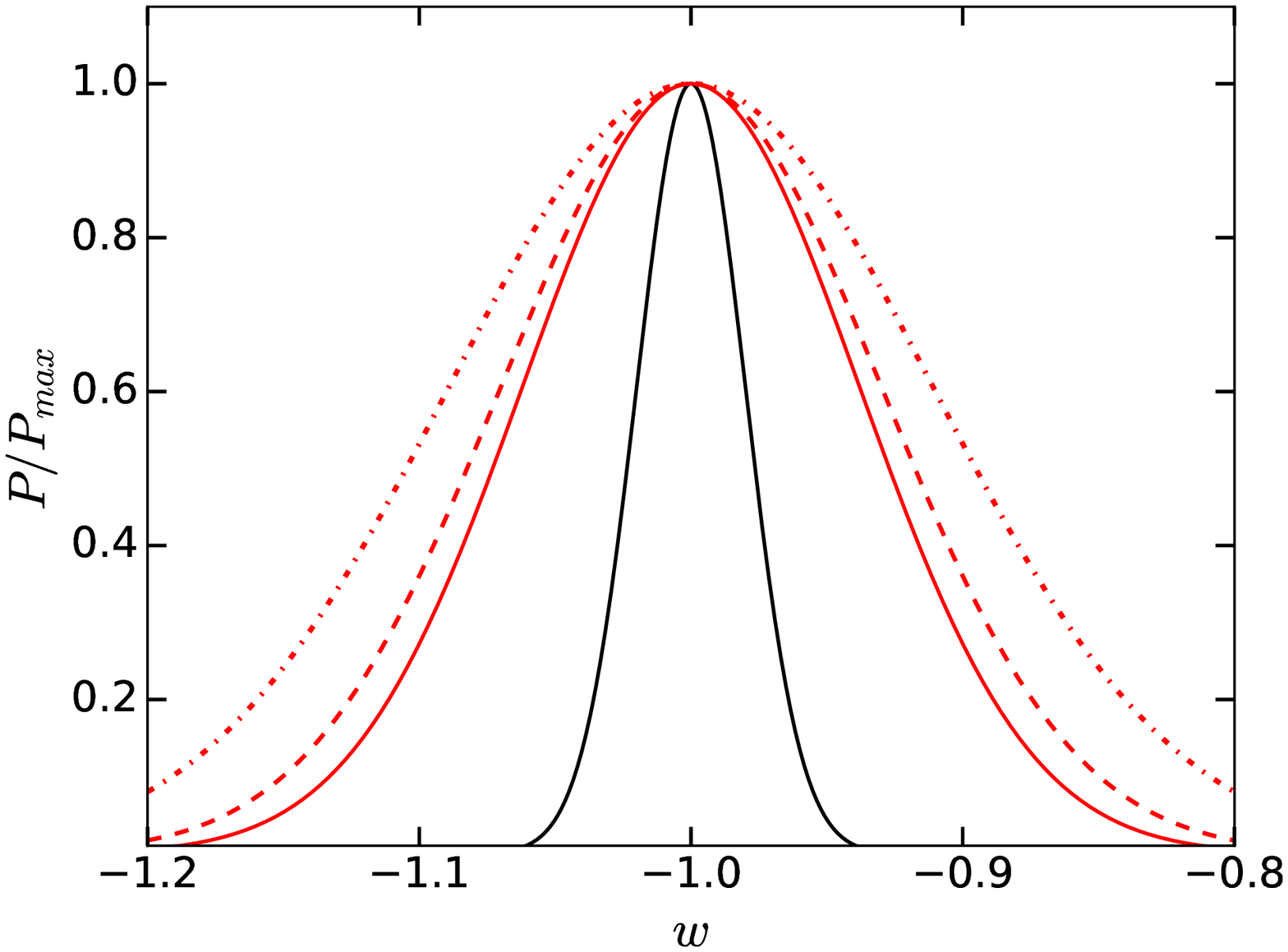}\\
     %\end{tabular}
   \caption{Marginalized probability distributions for the sum of neutrino masses, and dark energy equation of state $w$ from CMB Stage 4 lensing (red curves) compared to LSST (black curve). The solid red curve corresponds to $f_{\rm sky} = 0.75$, the dashed red curve corresponds to $f_{\rm sky} = 0.5$ and the dot-dashed curve corresponds to $f_{\rm sky} = 0.25$ for CMB Stage 4 experiments. While neutrino mass constraints from the two experiments are similar, $w$ can be constrained more tightly by the LSST survey.
   }
   \label{1dprob_cmb}
   \end{figure}

  Next, we compare the constraints on the parameters obtained from weak lensing in galaxy surveys to the constraints derived from CMB Stage 4 lensing, with the inclusion of {\it Planck} priors. 
  This comparison is illustrated in Fig. \ref{1dprob_cmb}. The black solid line represents the results from LSST, while the red lines represent the results from CMB Stage 4 experiments, for different sky coverages.
  We find that the constraints on the sum of neutrino masses from the two experiments are very similar, when we consider the most optimistic scenario for CMB Stage 4 lensing ($f_{\rm sky} = 0.75$), where the latter gives a constraint of $\sigma \big(\sum m_\nu\big)=0.046\,$eV. 
  
  %{\bf \textcolor{red} {In this case (with $N_{\rm eff}$ and no BAO priors), choosing $l_{\rm max} = 1000$ and $f_{\rm sky} = 0.5$ for both LSST and CMB Stage 4 lensing, the neutrino mass bound is $0.037\,$eV for LSST and $0.055\,$eV for CMB Stage 4. Looks like we don't gain much by just increasing $l_{\rm max}$ for LSST without changing the number densities - the noise starts dominating at $l \sim 400$ as seen in Fig. 2.}}
  It is interesting to compare the sensitivity of a CMB lensing survey and a weak lensing photometric to the sum of neutrino masses under the assumption that they cover the same fraction of the sky, and that they are sample variance dominated out to a similar range in multipole $l$. As can be seen from Fig. \ref{surveynoise} and Fig. \ref{cmbnoise}, CMB lensing is sample variance dominated out to $l\sim 1000$, whereas with the number densities of lenses and sources assumed for LSST, the signal is sample variance dominated out to $l\sim 400$. For this comparison, therefore, we tune the number densities of lenses and sources in LSST to 4 times their fiducial value, and then compare the constraints on $\sum m_\nu$ from the photometric galaxy survey and CMB lensing using $l_{\rm max} = 1000$ and $f_{\rm sky} = 0.5$ in both cases. Under these assumptions, we find $\sigma(\sum m_\nu) = 0.018\,$eV for the photometric survey and $\sigma(\sum m_\nu) = 0.055\,$eV for CMB lensing. This shows that the tomographic information in a photometric survey allows us to sample more modes, and therefore have a higher Signal to Noise Ratio (SNR) compared to a CMB lensing survey where the source redshift and the lensing kernel is fixed by the CMB last scattering surface. Quantitative comparisons of the SNR for a photometric survey like LSST and CMB lensing has been performed in \cite{Schaan2016}.

  For CMB Stage 4 lensing, we find that our constraint on $\sum m_\nu$ is weaker by about a factor of 2 compared to those in \cite{S4ScienceBook}. There are two main reasons behind this difference. First, \cite{S4ScienceBook} assumes priors on other cosmological parameters coming from DESI BAO measurements, which will be much tighter than the {\it Planck} priors that we use in this work. Secondly, $N_{\rm eff}$, which has large degeneracies with the neutrino mass is included in our analysis, while the constraints quoted in \cite{S4ScienceBook} assume that extremely accurate measurements of the CMB primaries from the Stage 4 experiments will constrain $N_{\rm eff}$ independently. Refs. \cite{Wu2014,Allison2015,Abazajian2013} also find very similar constraints on the neutrino mass when using DESI BAO priors in addition to CMB S4 lensing.
  
  To check how our estimates improve with the inclusion of low redshift information coming from DESI, we use the parameter forecasts provided in \cite{DESI} in our analysis. This extra information is especially helpful in tightening the constraint on the parameters $\Omega_m$ and $w$, coming from precise measurements of the BAO feature at low redshifts. We find that the 1-$\sigma$ constraint on the sum the of the neutrino masses from LSST clustering and lensing is $0.032\,$eV when information from DESI is included, compared to our fiducial result of $0.041\,$eV. For CMB Stage 4 lensing, the improvement is even more marked. With the extra constraints from DESI, Stage 4 CMB lensing can constrain the sum of neutrino masses to $0.029\,$eV at the 1-$\sigma$ level. The larger improvement in the CMB S4 bound is understandable since the CMB lensing kernel is peaked near $z\sim 2$, and therefore the information from low redshift provided by DESI is mostly complementary to the information contained in CMB lensing. On the other hand, for LSST, some of the information about low redshifts is already included as the lens bins in our analysis extend down to $z=0.4$. So adding in the DESI priors to LSST analysis do not improve the existing bounds on the neutrino mass by a lot.
  
  Next, we add additional priors on the parameter $N_{\rm eff}$ based on the forecasts for CMB Stage 4 experiments. We assume that $\Delta N_{\rm eff}$ will be measured using the CMB primaries at a 1-$\sigma$ error level of $0.03$. When this prior is included in our calculation for the LSST clustering and lensing, the 1-$\sigma$ constraint on the sum of neutrino masses tightens further to $0.028\,$eV. Similarly, when this prior on $N_{\rm eff}$ is included in the CMB Stage 4 lensing estimates, the constraint on the neutrino mass becomes $0.023\,$eV. These numbers are then comparable to the constraints on the neutrino mass forecast in Ref. \cite{S4ScienceBook}. 
  
  When compared to the current constraints on the neutrino mass coming from CMB lensing in the {\it Planck} experiment \citep{Planck2015}, we find that both LSST weak lensing and CMB Stage 4 lensing will improve the bounds on the neutrino mass by almost an order of magnitude. This is especially true once low redshift information from a DESI-like experiment is included for the CMB lensing analysis. For the {\it Planck} value, we have used the constraint obtained from the CMB lensing from {\it Planck} itself combined with {\it Planck} primaries only, with no additional external datasets. The 1-$\sigma$ constraint on the neutrino mass from this dataset only is $\sim 0.29\,$eV.
  
  Since LSST weak lensing and CMB Stage 4 lensing are sensitive to very different systematics, the fact that they are expected to provide very similar constraints on the sum of neutrino masses is an important result. Since the errors from the two experiments do not correlate, the statistical significance of a detection from one of experiments will be greatly enhanced when combined with the data from the other.
  
  While LSST weak lensing and CMB stage 4 lensing provide similar constraints on the sum of the neutrino mass, the constraint on $w$ from LSST is much stronger than the constraint from CMB Stage 4, even when we consider the most optimistic case for the latter. The CMB lensing kernel peaks at around $z\sim 2$, when dark energy forms a negligible fraction of the energy budget. On the other hand, LSST weak lensing is sensitive to much lower redshifts when dark energy starts to dominate. Further, the equation of state for dark energy affects the growth rate of the power spectrum. Since we use spectra from multiple redshift bins in LSST, the growth rate, and in turn, $w$, are better constrained than in CMB lensing.

%\begin{figure}
%\begin{tabular}{ccc}
%\centering
%  \includegraphics[scale = 0.45]{cmbnuw.eps}   \\
%     %\end{tabular}
%   \caption{1-$\sigma$ (solid lines) and 2-$\sigma$ (dashed lines) confidence intervals on 2-d subspace of the parameters $\sum m_\nu$ and $w$ from Stage 4 CMB lensing.}
%   \label{2dprob_cmb}
%   \end{figure}

%   \begin{figure}
% \centering
% \includegraphics[scale = 0.4]{nuprob_combined.eps}
% \caption{Comparison of the marginalized probability distribution of the sum of the neutrino masses for different datasets. The solid black line represents the LSST constraints with 6 source and lens bins. The dashed black line represents the most optimistic constraints expected from CMB Stage 4 lensing, while the dash-dotted black curve is for current constraints from Planck using CMB primaries and Planck lensing data.}
% \label{cmbcomparison}
%\end{figure}

   %  \subsection{Effect of systematics}
 %  \label{const_syst}

\subsection{Combination of LSST and CMB S4}
   It is also possible to combine measurements from LSST and CMB Stage 4 lensing assuming that the survey windows of the two overlap each other. We use data from these overlapping surveys to measure the cross-spectra $C_l^{g\kappa_{\rm CMB}}$, along with the galaxy autospcetra $C_l^{gg}$ from LSST and the shear-shear autospectrum $C_l^{\kappa_{\rm CMB}\kappa_{\rm CMB}}$ from CMB lensing. Using CMB lensing $\kappa_{\rm CMB}$ measurements reduces some of the systematic uncertainties, since the multiplicative shear bias and source redshifts uncertainties present in the LSST lensing measurements no longer affect the results.
   
\begin{figure}
%\begin{tabular}{ccc}
\centering
  \includegraphics[scale = 0.35]{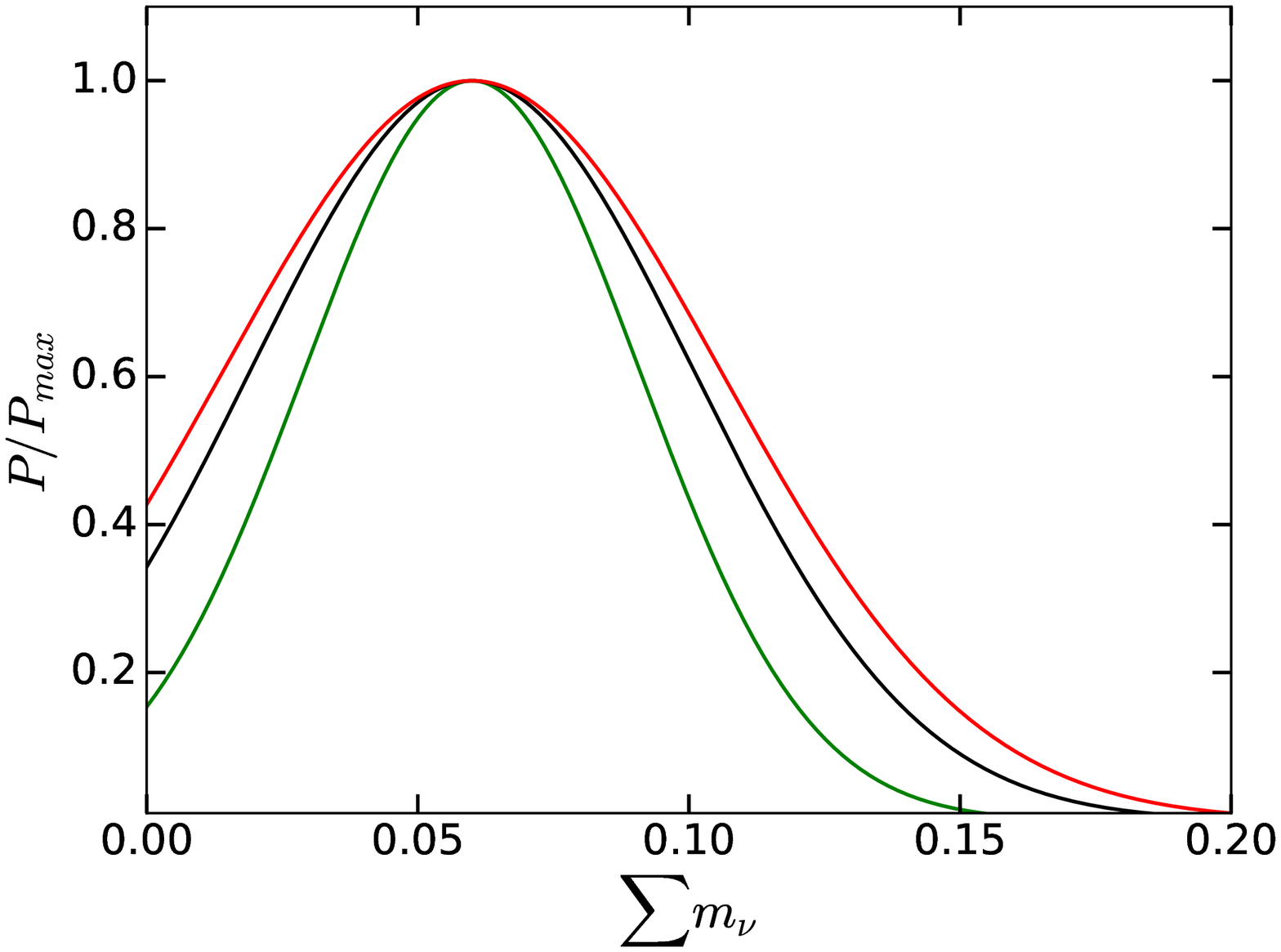}     \includegraphics[scale = 0.35]{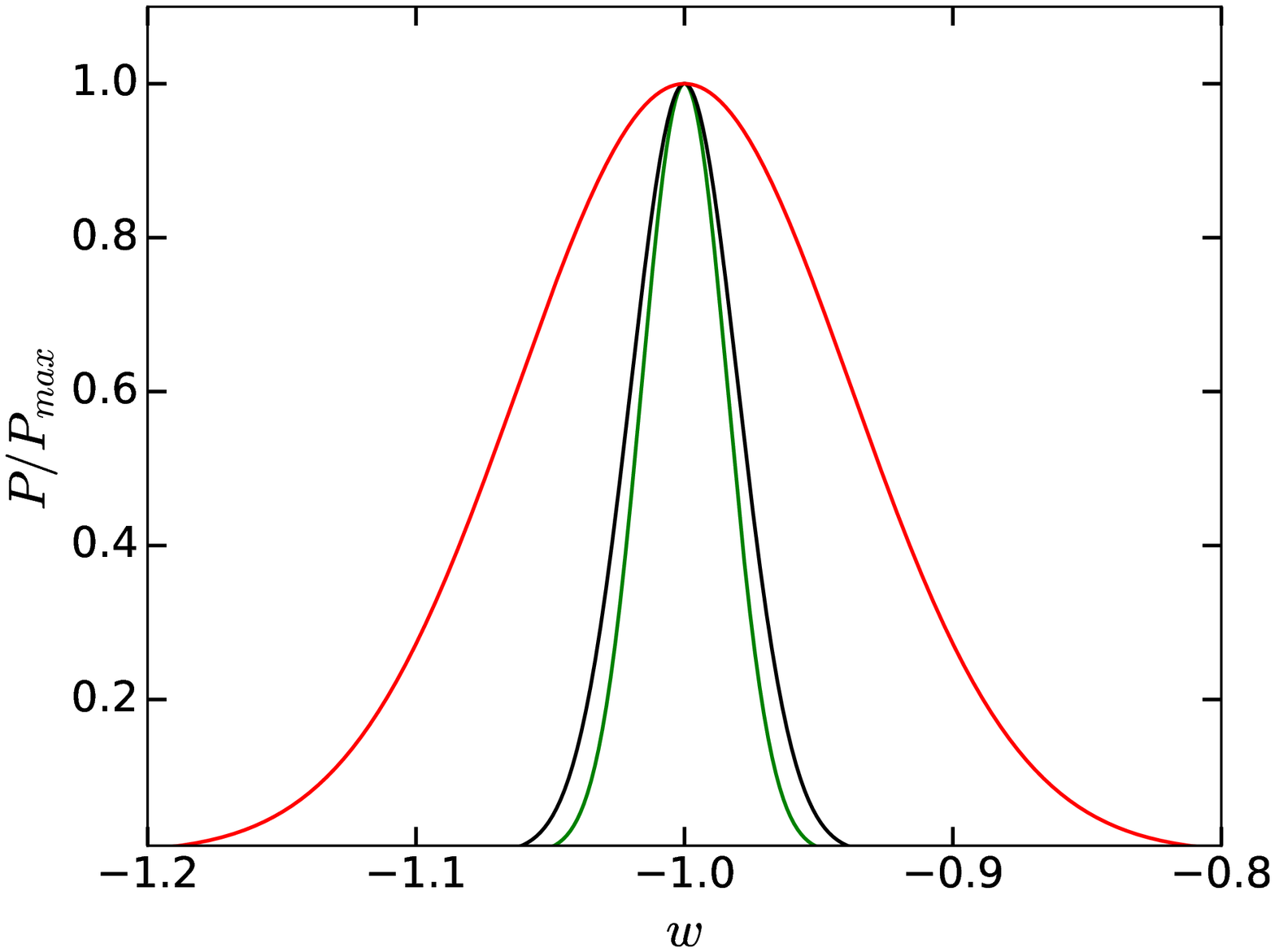}\\
     %\end{tabular}
   \caption{Marginalized probability distributions for the sum of neutrino masses $\sum m_\nu$, and dark energy equation of state $w$ for joint LSST and CMB lensing analysis (green curve) compared to those coming from LSST only (black curve) and CMB lensing only (red curve). The constraints improve somewhat marginally compared to the LSST only results.
   }
   \label{1dprob_kappacmb}
   \end{figure}

   For this analysis, we use the same lens redshift bins for LSST, as well as the same experimental sensitivity for  as mentioned in \S \ref{systematics}. We assume $f_{\rm sky} = 0.5$ for both experiments with complete overlap of survey windows. We find that this combination provides 1-$\sigma$ constraints of $\sigma\big(\sum m_\nu\big) = 0.031\,$eV and $\sigma(w) = 0.016$. Both these constraints are slightly stronger than the ones obtained from LSST only and the most optimistic case for CMB Stage 4 lensing, as shown in Fig. \ref{1dprob_kappacmb}. 
   
   One can also check how this joint constraint on the neutrino mass coming from LSST and CMB lensing improves when priors from the DESI experiment is included. Once again, we account for the extra information from DESI by adding stronger priors, especially on $\Omega_m$ and $w$, whose values are given in \cite{DESI}. We find that including these stronger priors improves the 1-$\sigma$ constraint on the sum of neutrino masses to $0.020\,$eV. This bound is competitive with the bounds presented in \cite{S4ScienceBook}, even though in our analysis, we marginalize over one extra cosmological parameter in $N_{\rm eff}$. On the other hand, this estimate may be a little optimistic given that we have assumed complete overlap of the survey volume between the LSST and CMB Stage 4 experiment.
   
   While the constraints on the cosmological parameters do improve when shear measurements from  CMB stage 4 are used, the improvement in the constraints is not dramatic unless extra information from a survey like DESI is included. On the other hand, adding CMB lensing to LSST helps with systematics of the LSST lensing measurements, especially in terms of constraining the multiplicative shear biases $m_i$ of the different redshift bins. This has been studied in detail in Ref. \cite{Schaan2016}. Therefore, even for cases where the overlap of survey volumes of the experiments is not perfect, useful information can be obtained by looking at the cross-correlations of observables from the two.
   
   \section{Dark radiation}
   \label{darkrad}
  Thermal dark radiation is a natural and generic prediction of many theories
beyond the standard model.  For instance, pseudo-Nambu-Goldstone bosons (pNGBs) are naturally light and are realized
in many extensions of the SM, arising from the spontaneous breaking of
a Peccei-Quinn symmetry (axions) \cite{Weinberg:1977ma,
  Wilczek:1977pj}, lepton number symmetry (majorons)
\cite{Chikashige:1980ui, Gelmini:1980re}, family symmetry (familons)
\cite{Wilczek:1982rv}, or dark number symmetry
\cite{Weinberg:2013kea}.  
%Perhaps the best motivated of these pNGBs
%are QCD axions.  Axions with $f_a \gtrsim 10^9$ GeV, allowed by bounds
%from energy loss in Supernova 1987A, can thermalize in the early
%universe \cite{Turner:1986tb, Masso:2002np} at temperatures $\gtrsim
%m_Z$ \cite{Salvio:2013iaa}, yielding a thermal relic population of
%$m_a \lesssim 6$ meV axions.  This population is out of reach of
%current cosmological data, thanks to the axion's small mass and high
%decoupling temperature, but represents a natural and potentially
%achievable target for the next generation of experiments
%\cite{Brust:2013xpv, Archidiacono:2015mda, Baumann:2016wac}. 
Given suitably large reheating temperatures and low symmetry breaking scales, these pGNBs can be thermally populated in the early universe.
Beyond these minimal models, string compactifications often yield a
proliferation of axion-like particles with a much expanded range of
masses and interaction strengths, making pNGBs a natural and generic
source of thermal bosonic dark radiation \cite{Svrcek:2006yi,
  Arvanitaki:2009fg,Jaeckel:2010ni}.  Such stringy
constructions can also yield dark $U(1)$ gauge bosons
\cite{Goodsell:2009xc, Cicoli:2011yh} with sub-eV masses, and possibly
also fermionic dark radiation in the form of photini
\cite{Arvanitaki:2009hb}.

Perhaps the leading candidate for thermal fermionic dark radiation is
a sterile neutrino.  
%Cosmological and terrestrial experiments together
%require that the production of light (sub-eV) sterile neutrinos
%through active-sterile mixing alone should yield a non-thermal relic
%population \cite{Hannestad:2012ky, Gariazzo:2015rra}\footnote{The
%  eV-scale sterile neutrinos favored by short baseline anomalies do
%  thermalize, resulting in contributions to $\Delta N_{\mathrm{eff}}$
%  at levels that are difficult to reconcile with Planck
%  measurements 
%  \cite{Mirizzi:2013gnd}. }. 
One way to thermalize sterile neutrinos is to add new interactions in the neutrino sector that are motivated by unification and by measurements of neutrino mixing
\cite{Engelhardt:2010dx, Anchordoqui:2011nh,
  Dasgupta:2013zpn,Ko:2014bka}. Similar to NNaturalness, thermal relic populations of both sterile neutrinos and dark photons can also arise in {\it mirror
sectors}, where the matter content of the SM is replicated, wholly or in
part, in a hidden sector.  This long-standing idea, reviewed in
\cite{Foot:2014mia}, has been motivated by parity restoration as well
as asymmetric dark matter and solutions to the hierarchy problem \cite{Arkani-Hamed2016,Chacko:2005}.  
%A recent resurgence of interest in mirror
%sectors has been driven by ``neutral naturalness'' solutions to the
%hierarchy problem, where mirror particles, rather than particles
%charged under SM gauge symmetries, are responsible for taming the
%quadratic divergence in the Higgs mass parameter \cite{Chacko:2005pe}
%\JS{fill in a few more of the recent refs}.
%These models have a hidden sector that only
%partially mirrors the SM, and thus may have thermal
%relativistic relics in the form of mirror neutrinos, massless {\it or}
%massive mirror photons, neither, or both. 
%\JS{add twin higgs cosmo refs}
 
More generally, hidden sectors constitute a generic possibility for
physics beyond the standard model.  Cosmologically, thermal dark
sectors are very well motivated as a source of DM.  
%In such models,
%the relic abundance of DM is determined by dark dynamics with little
%to no direct involvement of the SM, and DM-SM couplings may be
%parametrically small, thereby simply and economically explaining the
%lack of DM signals in direct detection and collider experiments to
%date.  
Such hidden sector models may be thermal,
e.g.~\cite{Pospelov:2007mp, ArkaniHamed:2008qn}, or nonthermal,
e.g.~\cite{Zurek:2013wia}.  In either case, the entropy of the hidden
sector must generically be either deposited into the SM thermal plasma
or carried by dark radiation.  Dark radiation is thus a
generic component of dark sector model-building.  The ratio of the
dark radiation temperature to the temperature of the SM will generally
depend on the physics of reheating \cite{Adshead:2016xxj} as well as
on the degrees of freedom in both sectors and the strength of the
leading coupling(s) between them, and for our purposes can
be treated as a free parameter.  Models where a dark radiation species
is directly involved in the freezeout of DM are frequently motivated
by structure formation, and in such models the radiation does
not always free-stream.  However, in the general case, the lightest stable
state(s) in the dark sector may be well-described by a thermal
free-streaming state, as in the SM; e.g.  \cite{Feng:2011ik}.
   
   \begin{figure}
 \centering
 \includegraphics[scale = 0.5]{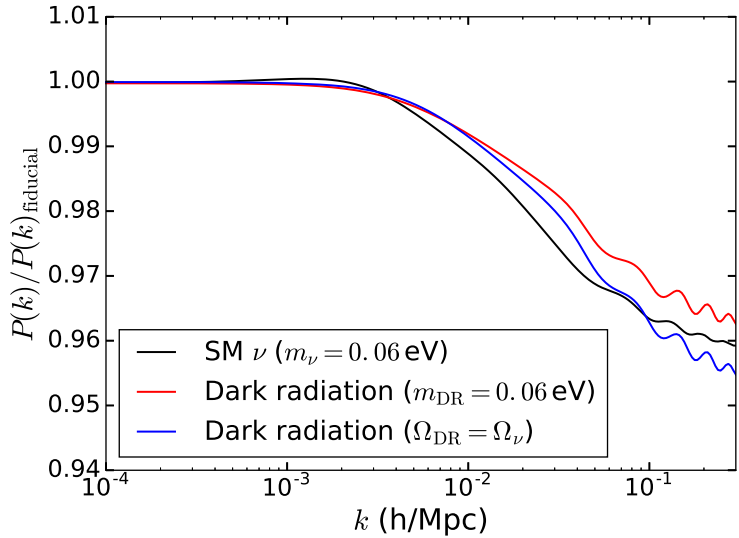}
 \caption{Damping in the power spectrum caused by a Standard Model neutrino (black curve) with mass $m_\nu = 0.06\,$eV compared to the damping due a fermionic dark radiation (red curve) particle producing $\Delta N_{\rm eff} = 0.15$ at CMB and having the same mass, $m_{\rm DR} = 0.06\,$eV. The shapes and the amplitudes of the damping are different for the two particles, implying that dark radiation and Standard Model neutrino masses can be measured simultaneously in surveys like LSST.}
 \label{fermioncomp}
\end{figure}

   An extra light degree of freedom should in principle show up in measurements of $N_{\rm eff}$ from the primary CMB, and the size of the signal is proportional to $T_{\rm DR,CMB}^4$, where $T_{\rm DR,CMB}$ is the temperature of the dark radiation species at the epoch of CMB last scattering.  {\it Planck} has already constrained $\Delta N_{\rm eff} < 0.33$, and future CMB Stage 4 experiments should be able to improve the constraints to $\Delta N_{\rm eff} \lesssim 0.03$ \cite{S4ScienceBook}.  However, it should also be noted that if the particles are relativistic at CMB last scattering, $\Delta N_{\rm eff}$ is insensitive to the mass of the dark radiation particle.  Apart from a signature on $N_{\rm eff}$, these light degrees of freedom can also damp the matter power spectrum at late times. The amount of damping is proportional to the energy density of the species today, $\Omega_{\rm DR}$. Assuming that these particles are non-relativistic at late times, the energy density is proportional to the mass of these dark radiation particles $m_{\rm DR}$, and to the late time number density $n_{\rm DR}$, i.e\ $\Omega_{\rm DR} \propto m_{\rm DR} n_{\rm DR}$.

   Since the damping of the matter power spectrum on small scales depends on $\Omega_{\rm DR}$, surveys like LSST will be sensitive to this parameter. We illustrate this with the following example, in which we suppose that $\Delta N_{\rm eff} = 0.15$ is observed in future CMB experiments.  We then repeat our Fisher matrix analysis, adding the mass of the dark radiation particle as an extra free parameter. The relevant power spectra and transfer functions were produced using the publicly available Boltzmann code CLASS \cite{CLASS1,CLASS2}, which allows for easy implementation of bosons as an extra light species. We find that for fermionic dark radiation, the constraint forecast is $\sigma(m_{\rm DR}) = 0.162\,$eV, and for bosonic dark radiation, $\sigma(m_{\rm DR}) = 0.137\,$eV. Note that the mass bound is somewhat different for fermions than for bosons, simply due to the difference in density arising from Fermi-Dirac or Bose-Einstein statistics \cite{Hannestad:2005}.  We further note that these results have been obtained after marginalizing over all other cosmological parameters including the neutrino mass. This means that LSST is potentially sensitive to multiple light species with different temperatures and masses, as the damping signatures will have different amplitudes and shapes, as illustrated in Fig. \ref{fermioncomp}.  The uncertainty on the inferred mass scales inversely with the magnitude of $\Delta N_{\rm eff}$ observed in the CMB.

   \section{NNaturalness}
   \label{NNat}
NNaturalness \cite{Arkani-Hamed2016} is a new approach to solving the Higgs hierarchy problem.    NNaturalness posits that there are $N$ copies of the Standard Model with differing Higgs masses.  The Higgs mass squared parameters are distributed between $-\Lambda^2$ and $\Lambda^2$, where $\Lambda$ is the scale which cuts off quadratic divergences.  If $N$ is large enough, one of the copies will naturally have a Higgs mass parametrically smaller than the cutoff. The sector with the smallest negative Higgs mass squared is identified with the Standard Model.

Given current constraints on $\Delta N_{\rm eff}$, the question of why the Higgs mass is small has been transformed into the question of why the sector with the lightest Higgs dominates the energy density of the universe. Crucially the lightest sector need to be dominantly reheated without making it otherwise special in any way.  Not abiding by this rule would reintroduce the hierarchy problem. 

To solve the problem, a new scalar field, the reheaton, was introduced.  If the reheaton couples with equal strength to the various copies of the Standard Model through the interaction $\phi H^\dagger_i H_i$ and is light (with a mass comparable to the lightest Higgs), then it predominantly reheats the sector with the lightest negative Higgs mass.  A light reheaton can only decay through off-shell Higgs bosons, which favors the lighter Higgs masses.  %Thus, the naturalness problem is solved.

The cosmological signals of this scenario come from the fact that the reheaton inevitably populates some of the other copies of the Standard Model. Since the sector with a light Higgs boson can not be singled out in any way, the new sectors are very similar to us. For the sake of calculability we assume that they are identical copies, except for slightly heavier Higgs bosons (with and without a vev). Relaxing this assumption does not appreciably change the phenomenology. 
The new copies are reheated to temperatures slightly smaller than our own and like our sector have many light particles (including neutrinos) and a massless photon.  The presence of these new light particles, with slightly lower temperatures than our own, makes this model an ideal candidate to be constrained and/or discovered by the techniques described before.

The cosmology of this realization of the NNaturalness paradigm is determined by two variables.  The first is the mass of the reheaton, $m_\phi$, and the second is the distribution of Higgs masses.  The coupling of the reheaton to the Higgs cancels out once the temperature of our sector has been fixed.  In principle, the differing Higgs masses can be drawn from any distribution. However any feature in the distribution would imply some assumptions on the dynamics related to the hierarchy problem at the scale $\Lambda$. So we take a uniform distribution with Higgs masses varying as
 \begin{equation}
  \left(m_H^2\right)_i = -\frac{\Lambda_H^2}{N} ( 2 i+ r ), \quad\quad  - \frac{N}{2} \leq i \leq \frac{N}{2}.
 \end{equation}
$r$ is a real and positive parameter that accounts for the possible probabilistic nature of the Higgs mass distribution or alternatively can be seen as a proxy for fine-tuning. If $r  = 1$ the Higgs masses are equally spaced around zero. For $r<1$ the lightest sector is closer to $m_h=0$. Thus the cosmology of this simple model of NNaturalness is determined by $m_\phi$ and $r$.

\begin{figure}
 \centering
 \includegraphics[scale = 0.5]{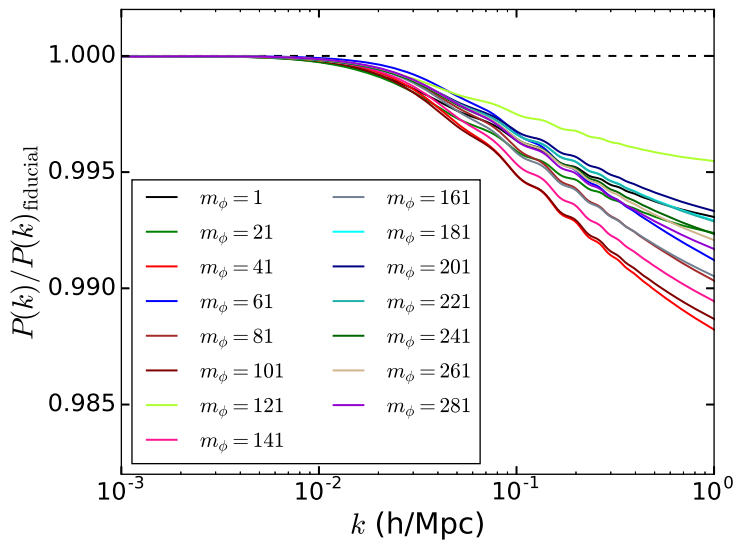}
 \caption{Ratio of the power spectrum at $z=0$ for different NNaturalness scenarios (labeled by $m_\phi$) compared to the Standard Model normal hierarchy scenario. The presence of extra free streaming species damps the power spectrum on the free streaming scale of the lightest of the extra species. On linear scales ($k>0.1\,$h/Mpc), the damping is below a percent level.}
 \label{NNatural_figure}
\end{figure}

In order to test the visibility of this model, we consider a set of different reheaton masses.  For each reheaton mass, we take the largest value of $r$ such that $\Delta N_\text{eff}$ is small enough to satisfy current constraints. This leads to different combinations of masses and temperatures for the extra neutrino species. We implement these extra neutrinos using the CLASS code \cite{CLASS1, CLASS2}.  The CLASS code  allows for extra non-CDM species with different masses, temperatures and number densities, and therefore it is straightforward to check the ratio of the 3-dimensional linear matter power spectrum for different extra neutrino masses and temperatures. The results are plotted in Fig. \ref{NNatural_figure}. The fiducial cosmology was taken to be  $\Lambda$CDM plus one massive neutrino species with $m_\nu = 0.06\,$eV. For each reheaton mass, we see the generic feature of extra damping on small scales, where the amplitude of the damping is proportional to the energy density in the extra copies of the neutrinos. The scale at which these damping effects start showing up is given by the free streaming scale of the extra neutrinos. Since the extra copies have a higher mass and a lower temperature than the Standard Model neutrino, in each case we find that the free streaming scale of the extra neutrinos is smaller than that of the Standard Model neutrino.

We then proceed to use the formalism from \S \ \ref{modeldiff} to check the statistical significance of the detection for these cosmologies with different reheaton masses. For the reheaton mass which produces the largest damping in the matter power spectrum, the calculated $\Delta \chi^2$ difference with the fiducial cosmology is $0.151$. 

This small $\Delta \chi^2$ suggests that these models will be hard to detect at LSST using information from linear scales only, along with the priors from {\it Planck}.  %For more optimistic forecasts, we can once again adopt a more aggressive approach with regard to the survey parameters, as well as our assumptions about linearity. As in \S \ref{darkrad}, we go up to $l_{\rm max} = 1000$ for all the lens and source bins, double the number densities of sources and lenses, as well as include priors from the upcoming DESI experiment - we believe all these possibilities are achievable by the time the LSST and CMB Stage 4 surveys are analyzed. 
To obtain more optimistic estimates, we modify some of our assumptions and survey parameters. First we set $l_{\rm max}$ for our various lens and source bins to $1000$. This obviously violates our assumptions of linear theory, but if nonlinear scales can be modeled correctly, information from these small scales could, in principle, enhance the detectability of the NNaturalness models. We clarify that in this analysis, we continue to use linear power spectra. This is likely a somewhat conservative estimate of the signal at small scales since nonlinear terms raise the amplitude of the power spectrum on these scales. It is worth noting that pushing to $l_{\rm max}\sim 1000$ at low redshifts also means that baryonic physics needs to be taken into account, which we can safely neglect at larger scales. This introduces a new systematic, and needs to be modeled accurately, either through simulations, or semi-analytics models to extract information about cosmological parameters from small scales, e.g.~\cite{Eifler2014,Zentner2013}.
The next assumption we make is to double the source and lens galaxy counts in the different bins. For the lens bins, the redMaGiC galaxies are only a fraction of all galaxies in the survey at that redshift, therefore doubling the number of galaxies for which the redshifts are known accurately seems an achievable target. Trying to double the number of sources seems a more optimistic assumption. Finally, we also add tighter priors on $\Omega_m$ and $w$ from BAO measurements in DESI. 
We repeat our analysis to calculate $\Delta \chi^2$ for the NNaturalness models with respect to the fiducial cosmology, and we find that even under these assumptions, the largest $\Delta \chi^2$ for the NNaturalness realizations considered here is $0.53$.

Since the combinations of $m_\phi$ and $r$ that were considered here produced the highest $\Delta N_{\rm eff}$ consistent with current bounds, measurements of $N_{\rm eff}$ from Stage 4 CMB experiments will easily detect the contribution from such models. But the detection will not be able to distinguish them from other scenarios e.g.~\cite{Dasgupta:2013zpn,Weinberg:2013kea}. We have explored the detection of its signature in in large-scale structure. We find that even the lensing surveys of the next decade will be hard pressed to make a detection. The signal is below a percent-level suppression of the power spectrum, which is challenging to achieve statistically and, even if we pushed survey parameter to achieve that, it would be daunting to control systematics at the desired level. One might hope that a completely different observable that goes to smaller scales or otherwise samples many more modes – like the Lyman-$\alpha$ forest or 21cm probes of structure - while achieving sub-percent control of systematics, will be sensitive to these cosmologies.

   \section{Summary and Discussion}
   
   \label{summary}
   
   We have obtained forecasts for the constraints on the neutrino mass from measurements of galaxy clustering and weak lensing in the LSST experiment, using information from only linear scales. The constraints on cosmological parameters, including the sum of the neutrino mass, improve with the number of redshift bins used in the analysis. We find that using
    6 lens redshift bins and 6 source redshift bins, the sum of neutrino masses can be constrained to $\sigma\big(\sum m_\nu\big) = 0.041\,$eV and the dark energy equation of state can be constrained to $\sigma(w)=0.020$. While the constraint on the neutrino mass is very similar to that coming from Stage 4 CMB lensing experiments, LSST provides much stronger constraints on $w$ than CMB lensing. These constraints were obtained by including priors on cosmological parameters coming from the {\it Planck} experiment. Further, the number of relativistic degrees of freedom $N_{\rm eff}$ was included as a parameter in our analysis. In principle, $N_{\rm eff}$ could be constrained independently from measurements of the CMB primaries in the CMB Stage 4 experiments. However, to provide a conservative estimate of the constraints, we initially include it as a parameter that is to marginalized over when obtaining constraints on other cosmological parameters of interest. Our constraints are in agreement with similar forecasts in literature e.g.~\cite{Hannestad2006,Kitching2008}.

    When we include the expected tighter priors on other cosmological parameters from the upcoming survey DESI, the forecasted 1-$\sigma$ constraint on the sum of neutrino masses from LSST goes down to $0.032\,$eV. A similar analysis for CMB Stage 4 lensing yields a constraint of $0.029\,$eV. These improvements come mainly from the tighter constraints on $\Omega_m$ and $w$, arising from accurate BAO measurements  at low redshifts.
    
    When we further included strong priors on $N_{\rm eff}$, assuming that it will be measured independently in upcoming surveys, the forecasted constraint on the sum of neutrino masses at LSST tightens to $0.028\,$eV. For CMB Stage 4 lensing, this additional prior on $N_{\rm eff}$ yields a constraint of $\sigma(\sum m_\nu) = 0.023\,$eV. These constraints on the sum of neutrino masses from CMB lensing are consistent to those presented in \cite{Allison2015} when marginalizing over the same set of cosmological parameters.
    
    We have also obtained constraints on $\sum m_\nu$ and $w$ using a combination of galaxy clustering measurements from LSST and CMB stage 4 lensing. Once again, if low redshift information from DESI is included in the analysis, the constraints on the sum of neutrino masses becomes tighter. In fact, this combination can constrain $\sigma(\sum m_\nu) = 0.020\,$eV, which is comparable to the bounds obtained from CMB lensing only in \cite{S4ScienceBook,Allison2015,Wu2014,Abazajian2013}, even though we marginalize over an extra parameter in $N_{\rm eff}$ for this comparison. Note that this combination of LSST, CMB Stage 4 lensing and DESI would be able to detect the minimal mass normal hierarchy of neutrinos at $\sim 3\sigma$. However, this is an relatively optimistic estimate. Firstly, this estimate assumed perfect overlap between the survey volumes of LSST and CMB Stage 4 experiment, and that the cross-correlations of data from the two surveys does not throw up unforeseen issues. This also assumes that all the major sources of systematics have been accounted for and behave as expected, and that all the surveys are able meet their target statistical error levels. On the other hand, it might be possible to extract even stronger constraints using this technique if smaller scales at LSST can be modeled more accurately.

   %One of the main sources of systematics in the LSST measurements will be shear bias and uncertainties in the photometric redshift of the source galaxies. We use RedMagic galaxies, for which the photometric redshifts are measured to a very high accuracy, for lens galaxies, and therefore neglect lens redshift errors as a source of systematic errors. Using the parameterization described in \S \ \ref{systematics}, we investigate how the constraints on $\sum m_\nu$ and $w$ change with different levels of prior on the nuisance parameters. We find that even without any prior on these extra parameters, the constraint on $\sum m_\nu$ degrades by $\sim 17\%$, while the degradation of the constraint on $w$ is even smaller. Using a reasonable prior of $0.02$ on each of the nuisance parameters degrades the above constraints only by a few percent. 
   
   One of the main sources of systematics in the LSST measurements will be shear bias and uncertainties in the photometric redshift of the source galaxies. We use redMaGiC galaxies, for which the photometric redshifts are measured to a very high accuracy, for lens galaxies, and therefore neglect lens redshift errors as a source of systematic errors. However, we account for redshift uncertainties of source galaxies and uncertainties in the shear bias using the parameterization in \S \ \ref{systematics}. Our constraints were obtained without placing any priors on the nuisance parameters $m_i$. We have also investigated how the constraints improve when priors are placed on these parameters. The improvement is quite weak in $\sum m_\nu$, and even weaker for $w$. Placing a realistic prior of $0.02$ on each of the $m_i$ produces an improvement of $\sim 15\%$ on the neutrino mass constraint, and $\sim 5\%$ improvement on the error on $w$.
   
   We note that we have made some simplifying assumptions in this study. We have assumed a flat cosmology in our analysis, and so $\Omega_k$ is fixed to $0$ for all our calculations. We have also assumed that the equation of state for dark energy is time invariant, setting $w_a=0$. The effect of marginalizing over $\Omega_k$ and $w_a$ as extra parameters on the neutrino mass constraint has been studied in e.g.~\cite{Allison2015} for Stage 4 CMB lensing. It should also be pointed out we do not consider the effects of extended theories of gravity on the power spectrum, which can be degenerate with the effect of massive neutrinos \cite{Bellomo2016}. In this analysis, we have obtained the constraints on the sum of neutrino masses with the minimal mass normal hierarchy as the fiducial model. Various authors have looked into distinguishing the normal hierarchy of neutrino masses from the inverted hierarchy using cosmological observables, \cite{Hamann2012,Verde2015,Giusarma2016,Zablocki2016,PDelabrouille2014} for example. Moving on to systematics, we have neglected the effect of intrinsic alignments of galaxies and other effects which as discussed above are expected to be subdominant to the uncertainties considered in our analysis.

   %We find that while LSST is able to provide tight constraints on the neutrino mass using information from linear scales only, it is not able to differentiate other models with extra light degrees of freedom with sufficient statistical significance. We specifically looked at two types of models which have extra weakly interacting light degrees of freedom - the Naturalness model and models of thermal dark radiation. In both cases, we find that LSST will be able to detect them at a statistical significance of $\sim 0.2 \sigma$.
   
   Beyond Standard Model neutrinos, thermal dark radiation can also lead to observable signatures on both $N_{\rm eff}$ at CMB, as well as on the late time matter power spectrum. Dark radiation candidates which were relativistic at CMB, but non-relativistic today damp the power spectrum on scales below their free streaming scale with an amplitude that is proportional to the mass of the particle. In this paper, we have studied the bounds that can be placed on the mass of dark radiation particles if they produce a detection of $\Delta N_{\rm eff} = 0.15$ in the CMB primaries. The constraint on the mass depends on whether this new thermal species follows a Bose-Einstein distribution or a Fermi-Dirac distribution. For the former, we find a constraint on the mass $\sigma(m_{\rm DR}) = 0.137\,$eV at a 1-$\sigma$ level. For the latter, the mass constraint is $\sigma(m_{\rm DR}) = 0.162\,$eV at the 1-$\sigma$ level. We find that the mass constraints scale inversely as $\Delta N_{\rm eff}$. Since the neutrino mass was one of the parameters that was marginalized over to obtain this estimate, it suggests that LSST will be potentially sensitive to multiple light species with different temperatures and masses. 
   
   %In this paper, we have studied the case of differentiating thermal bosonic dark radiation particle from thermal fermionic dark radiation particles even if they produce the same $\Delta N_{\rm eff}$ at CMB, and have the same mass. We find that for optimistic choices of temperature, mass and the relative number of degrees of freedom of the boson and the fermion, LSST will be able to differentiate between these two types of dark radiation.  In the scenario where the boson is a vector and the fermion is Majorana, and for our aggressive approach - using information from nonlinear scales, assuming larger source and lens densities, and including constraints on other cosmological parameters from BAO measurements in DESI - the statistical significance of the detection is $1.45\, \sigma$. It is below $1\,\sigma$ for our more conservative, linear regime analysis.
   
   A specific model that posits the presence of extra light degrees of freedom is the NNaturalness model outlined in \cite{Arkani-Hamed2016}. This model predicts the presence of multiple copies of neutrinos with different temperatures and number densities, all of which can contribute to $N_{\rm eff}$. We have analyzed how effective LSST will be at detecting these extra neutrino species.
   We find that even when the $\Delta N_{\rm eff}$ produced by these extra neutrinos is near the current bounds, LSST measurements will only be able to detect the effects of these at a statistical significance of $0.151\,\sigma$. Even a more optimistic estimate yields a maximum $\Delta \chi^2 = 0.53$ between the fiducial cosmology and a cosmology including extra neutrinos from the NNaturalness models. Since CMB Stage 4 will improve current constraints on $N_{\rm eff}$ by almost an order of magnitude, the contribution to $N_{\rm eff}$ from the NNaturalness model will be detectable. However, even for an unambiguous detection of $N_{\rm eff}$ from measurements of the CMB primaries, it will be difficult to distinguish NNaturalness from other exotic models which also predict extra light degrees of freedom during the CMB epoch e.g.~\cite{Dasgupta:2013zpn,Weinberg:2013kea}.

   %It is conceivable that when data from other experiments like DESI, which will provide much stronger constraints on other parameters like $\Omega_m$ and $w$, is folded into the analysis, the statistical significance of this detection could become significantly higher.
   
   Through most of our analysis, we have only used modes for which linear perturbation theory is still valid. This allowed us to use a linear matter power spectrum instead of the full nonlinear matter power spectrum, as well to use the simplifying assumption that the covariance of various $l$ modes are independent of each other. Moreover, at these linear scales, we could make the added simplifying assumption that galaxy bias is scale-independent.  However, smaller scales which are mildly nonlinear, potentially have a lot of useful information due to the fact that there are many such independent patches on the sky. Extracting this information could lead to stronger constraints on the neutrino mass, and statistically significant  detection of extra light degrees of freedom. To do this, we first need to accurately model the  
   matter power spectrum in the presence of massive neutrinos and other light degrees of freedom. Secondly, we need to take into account off-diagonal covariances between different $l$ modes in the nonlinear regime. And finally, to use the galaxy-galaxy autocorrelation and the galaxy-galaxy lensing spectra on small scales, we need an accurate model of galaxy bias on these scales, where it can be strongly scale dependent. Using this extra information will allow for even stronger constraints on the neutrino mass, as well a more statistically significant detections of other light degrees of freedom in the future.
   
   Our study of dark radiation models makes clear that surveys planned for the next decade have interesting new discovery space. Indeed, for models such as NNaturalness, even more ambitious surveys will be required. One possibility is 3-dimensional surveys that can access more modes of mass fluctuations than any of the surveys discussed in this paper.  Surveys discussed in the Cosmic Visions whitepaper \cite{Dodelson2016}, typically for the 2030’s, would achieve fractional errors on the power spectrum of well below a percent. Such surveys would enable high significance detections of standard model neutrino properties and also test a variety of models for new particle species.
   
%\red{acknowledge CAMB, CLASS, Nima, Anson, Raffaele, Mat, Emmanuel}
\acknowledgments

We thank Anson Hook, Raffaele Tito D'Agnolo and Nima Arkani-Hamed for helpful discussions of NNaturalness. We are very grateful to Vinicius Miranda and Masahiro Takada for numerous suggestions and related collaborative work. We thank Mathew Madhavacheril and Emmanuel Schaan for helpful comments on an earlier version of this paper. BJ is grateful to Juliana Kwan, Marilena LoVerde, Niall MacCrann, Blake Sherwin and the neutrino study group at Penn's Center for Particle Cosmology for helpful discussions.
We also thank the developers of the CAMB and CLASS codes, which have been used extensively in this work. BJ is partially supported by the US Department of Energy grant DE- SC0007901. This work benefitted from discussions at the Aspen Center for Physics, which is supported by National Science Foundation grant PHY-1066293.
   %%%%%%%%%%%%%%%%%%%%%%%%%%%%%%%%%%%%%%%%%%%%%%%
   \bibliography{neutrino}
\bibliographystyle{JHEP}
\end{document}